\documentclass[11pt]{article}

\usepackage[margin=2.5cm]{geometry}
\usepackage{multicol}
\usepackage{amssymb}
\usepackage{float}
\usepackage{graphicx}
\usepackage[multidot]{grffile}
\usepackage{subcaption}
\usepackage[final]{pdfpages}
\usepackage{amssymb,amsmath}
\usepackage{mathtools} 
\usepackage{listings} 
\usepackage{enumitem} 
\usepackage[style=numeric,isbn=false,url=false,giveninits=true]{biblatex}
\usepackage[export]{adjustbox}
\DeclareFieldFormat[article,inproceedings]{titlecase}{\MakeSentenceCase*{#1}}
\usepackage{setspace} 
\usepackage{algorithm}
\usepackage[unicode=true,
  linktocpage,
  linkbordercolor={0.5 0.5 1},
  citebordercolor={0.5 1 0.5},
  colorlinks=true,
  linkcolor=blue]{hyperref}
\makeatletter
\graphicspath{{figure/}}

\DeclareMathOperator*{\argmin}{argmin}
\DeclareMathOperator*{\argmax}{argmax}

\newcommand{\pd}[2]{\frac{\partial #1}{\partial #2}}
\newcommand{\dv}[2]{\frac{\text{d} #1}{\text{d} #2}}

\newcommand{\Lag}{\mathcal{L}}

\newcommand{\Cmodel}{C_{\textrm{model}}}

\newcommand{\btheta}{\boldsymbol{\theta}}

\newcommand{\tth}{^{\text{th}}}

\newcommand{\bx}{\boldsymbol{x}}
\newcommand{\by}{\boldsymbol{y}}
\newcommand{\bu}{\boldsymbol{u}}
\newcommand{\bff}{\boldsymbol{f}}
\newcommand{\bxi}{\boldsymbol{\xi}}
\newcommand{\balpha}{\boldsymbol{\alpha}}
\newcommand{\blambda}{\boldsymbol{\lambda}}
\newcommand{\reff}{r_{\textrm{eff}}}
\newcommand{\keff}{K_{\textrm{eff}}}
\newcommand{\umax}{u_{\textrm{max}}}
\newcommand{\bunew}{\boldsymbol{u}^{\textrm{new}}}

\begin{document}

\title{Optimal experiment design for practical parameter identifiability and model discrimination}

\author{
Yue Liu$^{1,2*}$, Philip K. Maini$^{1}$, Ruth E. Baker$^{1}$ \\
$^{1}$Mathematical Institute, University of Oxford\\
$^{2}$Department of Mathematics, Purdue University\\
$^*$ Corresponding author: liu4194@purdue.edu
}
\date{}


\maketitle


\begin{abstract}
Mechanistic mathematical models of biological systems usually contain a number of unknown parameters whose values need to be estimated from available experimental data in order for the models to be validated and used to make quantitative predictions. This requires that the models are practically identifiable, that is, the values of the parameters can be confidently determined, given available data. A well-designed experiment can produce data that are much more informative for the purpose of inferring parameter values than a poorly designed experiment. It is, therefore, of great interest to optimally design experiments such that the resulting data maximise the practical identifiability of a chosen model. Experimental design is also useful for model discrimination, where we seek to distinguish between multiple distinct, competing models of the same biological system in order to determine which model better reveals insight into the underlying biological mechanisms. In many cases, an external stimulus can be used as a control input to probe the behaviour of the system. In this paper, we will explore techniques for optimally designing such a control for a given experiment, in order to maximise parameter identifiability and model discrimination, and demonstrate these techniques in the context of commonly applied ordinary differential equation models. We use a profile likelihood-based approach to assess parameter identifiability. We  then show how the problem of optimal experimental design for model discrimination can be formulated as an optimal control problem, which can be solved efficiently by applying Pontryagin's Maximum Principle.


\end{abstract}


\section{Introduction}

Mechanistic mathematical models are widely used to describe the dynamics of a biological system, which makes them invaluable for understanding the underlying mechanisms driving the behaviours of the system. Such models usually contain a number of parameters, whose values are often difficult to measure directly, and therefore need to be estimated by calibrating the model to available experimental data. Certain experimental data can be more informative than others for the purpose of inferring model parameters. This naturally raises the question of how to optimally design an experiment such that the data it produces are the most informative. This idea dates back over a century, as discussed in~\cite{kreutz2009SystemsBiologyExperimental}, with~\cite{hunter1965DesignsDiscriminatingTwo} serving as an early example. In mathematical biology, the ideas of experimental design have been applied to models of signalling pathways~\cite{bandara2009OptimalExperimentalDesign,cho2003ExperimentalDesignSystems}, gene regulatory networks~\cite{steiert2012ExperimentalDesignParameter}, yeast fermentation~\cite{espie1989OptimalDesignDynamic}, and PKPD models to study the effects of drugs~\cite{smucker2018OptimalExperimentalDesign}, and also in physics to models of crystallisation~\cite{chen2003DesignOptimallyInformative} and polymerisation~\cite{burkeannettel1994ModelDiscriminationDesigned}. These studies illustrate that a few well-designed experiments have the potential to provide as much, or even more, insight into the system being modelled as many naively-designed ones. Therefore, optimal experimental design is of great interest, as it can reduce costs by requiring fewer experiments, and improves the effectiveness of experiments in biological research.

In this paper, we discuss optimal experimental design in the context of ordinary differential equation (ODE) models, which are one of the most common types of mechanistic models in mathematical biology. Two important tasks in a modelling study are parameter estimation and model selection. These two objectives are inherently intertwined, and can both benefit from careful experimental design. The ability to accurately estimate parameter values is referred to as practical parameter identifiability, and can be understood as the inverse of the uncertainty in parameter estimates. Therefore, designing an experiment to optimise parameter identifiability aids parameter estimation. Alternatively, we can optimise the experiment for model discrimination, aiming to maximise the discrepancy between the predictions made by competing models: this discrepancy in predictions makes it easier to distinguish between the models given data, and to select the model that most accurately represents the biological system. Although these objectives share a connection, they are nonetheless distinct. Previous authors recognised this connection, as reflected in numerous papers that discuss both simultaneously~\cite{espie1989OptimalDesignDynamic,kreutz2009SystemsBiologyExperimental}. However, it is worthwhile to elaborate on the similarities and distinctions between them.

\subsection{Parameter identifiability}
\label{sec:intro_iden}

Practical parameter identifiability refers to the ability to confidently identify the values of model parameters given available data, which is often noisy, and limited in resolution~\cite{wieland2021StructuralPracticalIdentifiability}. Therefore, practical identifiability is a property of the combination of the model and the dataset. Practical identifiability should be distinguished from structural identifiability, which refers to the ability to uniquely determine parameter values in an idealised situation where the data are free from observational noise, and of arbitrarily high precision~\cite{bellman1970StructuralIdentifiability,cobelli1980ParameterStructuralIdentifiability}. 
Therefore, practical identifiability is a strictly stronger condition than structural identifiability. We focus on practical, rather than structural, identifiability in this paper, since practical identifiability accounts for the noisy and limited nature of realistic experimental datasets.

Designing an experiment for optimal practical identifiability revolves around considering a single model, where parameter values are initially unknown. A well-designed experiment will provide sufficient data to constrain the possible values of these parameters. The objective of experimental design, then, is to design an experiment that reduces the uncertainty in parameter estimates to the greatest extent. In order to evaluate the performance of an experimental design, we must have a way to measure parameter identifiability. Identifiability analysis is inherently tied to parameter inference, and each method of parameter inference gives rise to a different way to measure practical identifiability.

First, the frequentist maximum likelihood approach gives rise to the Fisher Information Matrix (FIM), which reflects the local curvature of the likelihood function at the Maximum Likeliood Estimator (MLE), with larger eigenvalues implying a sharper likelihood function, hence less uncertainty in parameter values. Based on this, some reasonable design objectives are to maximise the trace of the FIM, or the determinant, or the smallest eigenvalue. These are referred to as the A-, D-, and E-optimal criteria, respectively~\cite{faller2003SimulationMethodsOptimal,kreutz2009SystemsBiologyExperimental}. 
Second, in the framework of Bayesian inference, the variance and entropy (and its derived variants) of the posterior distribution of parameter values are all reasonable measures of uncertainty, as used in~\cite{vanlier2012BayesianApproachTargeted}.
Thirdly, under a profile likelihood-based approach, we can use the widths of the confidence intervals for parameter values (with a fixed confidence threshold, usually 95\%) as a measure of uncertainty. More complicated measures, such as the one defined in~\cite{litwin2022OptimalExperimentalDesign}, avoid the need to choose a confidence threshold, albeit having a much higher computational cost. The use of profile likelihoods for experimental design has been examined in~\cite{steiert2012ExperimentalDesignParameter}, but its use is less widespread than the other two approaches.

In this paper, we will employ the profile likelihood approach, with the width of the 95\% confidence interval serving as the measure of uncertainty to be minimised. Profile likelihoods strike a good balance between providing information about the parameters and computational cost, as argued in~\cite{liu2024ParameterIdentifiabilityModel,villaverde2022AssessmentPredictionUncertainty}.
The FIM-based metrics of identifiability, which are the most widely used metrics found in the literature, rely on point estimates and implicitly presume that the log-likelihood function approximates a parabolic shape (or, in higher dimensions, an ellipsoid), at least near the MLE. This assumption does not hold when the parameters are non-identifiable, making the FIM an inaccurate measure of parameter uncertainty~\cite{villaverde2022AssessmentPredictionUncertainty}. In particular, in the structurally non-identifiable case, the FIM is singular, which means the method cannot be directly used without adaptation. Methods based on Bayesian inference provide a much more accurate representation of the distribution of parameter values, but typically require a large number of model simulations, therefore incurring a high computational cost. The profile likelihood approach compromises between the two. In this paper, we will demonstrate how to design experiments to maximise parameter identifiability, as measured with the widths of the univariate 95\% confidence intervals constructed using profile likelihoods, for ordinary differential equation (ODE) models arising from three different biological applications.

\subsection{Model discrimination}
\label{sec:intro_discrim}

In contrast to the problem of improving parameter identifiability, model discrimination involves two or more models, which we assume contain no undetermined parameters. In the context of model selection, these models might represent alternative mechanistic explanations for the same biological observations, where the available data from previous experiments are insufficient for determining which one is more appropriate. In some cases, these models can involve different mechanisms or components, composed of distinct equations. In other cases, these models might share the same equations, except with different values for the coefficients in these equations. In order to select the most appropriate model, we need to design a new experiment in such a way that the predictions of the models differ as much as possible.

The notions of model discrimination and parameter identifiability are distinct, but related. Consider a scenario where the practical identifiability of a model with undetermined parameters is poor given existing data, and model fitting yields multiple parameter sets that reproduce the data well and represent distinct hypotheses about the underlying biological mechanisms. We can recast the problem of overcoming non-identifiability as a model discrimination problem, by selecting two or more parameter sets, and design a new experiment to maximise the difference in model output under the selected parameter sets. For example, in the case of a multi-modal profile likelihood, we can choose parameter sets at the local maxima of the profile likelihood. In cases where the confidence regions for the parameter values are very wide, we can choose parameter sets that correspond to the MLE, and others near the edge of the confidence region. In this way, we can iteratively design experiments to provide additional data that improves the identifiability of a model. Such an iterative approach is common in biological research~\cite{faller2003SimulationMethodsOptimal,gadkar2005IterativeApproachModel,kreutz2009SystemsBiologyExperimental}. The approach described in Sec.~\ref{sec:intro_iden} would be more suitable for the first iteration of model parameterisation, while the model discrimination approach for experimental design would be more suitable for later iterations, since it assumes more knowledge of the parameter values that the other approach.

\subsection{Design of experimental inputs}

Now that we are clear about the objectives of the experimental design, we ask the question of what aspects of the experiment can be changed for optimisation? For a dynamical model in mathematical biology, these aspects include the initial conditions of the experiment, the time span, the measurement or observation times and frequency, the observation procedure, any external stimuli or input to the experiment, and more. Each of these has received some attention in the literature~\cite{kreutz2009SystemsBiologyExperimental,raue2010IdentifiabilityObservabilityAnalysis,wieland2021StructuralPracticalIdentifiability}. Deciding on what quantities to measure is also an important part of experimental design. In many cases, it can be difficult and costly to measure certain variables in a model, therefore not all variables are measured. For example, in a model for glucose-insulin interaction~\cite{chin2011StructuralIdentifiabilityIndistinguishability}, insulin excitable tissue is usually not measured during experiments due to technical constraints. Such ``missing" measurements can lead to both practical and structural identifiability issues, since we are not then able to use the trajectories of the unobservable quantities to constrain parameter values.

Given the many possible aspects of experimental design, we are not able to discuss all of them in this paper. We therefore choose to focus on optimally designing an external stimulus, or control input, to the experiment. In existing works that discuss the design of controls for experiments, the control input is often restricted to take simple forms. For example, Bandara et al.~\cite{bandara2009OptimalExperimentalDesign} restricted the control to be step functions, Roberts et al~\cite{roberts2009ModelInvalidationbasedApproach} and Hamadeh et al~\cite{hamadeh2011FeedbackControlArchitecture} focused on sinusoidal functions, and Faller et al.~\cite{faller2003SimulationMethodsOptimal} used low degree polynomials---all of these approaches limit the degrees of freedom for designing the control. While these restrictions can be reasonable in experiments where the ability to precisely manipulate the stimulation is limited, the availability of modern instruments that allow for much more precise control over elements of the experiment means it is appropriate to relax some of these restrictions, and consider a more general class of controls. In this paper, we consider both continuous control, where the control can be arbitrary piecewise continuous functions of time, and ``bang-bang" control~\cite{bellman1956BangbangControlProblem}, where the control is restricted to a small number of discrete values, usually two.

In order to constrain the design space, all other aspects of the experiment will be considered fixed, and cannot be changed in the design. Notably, we assume that the observations are made at evenly spaced times $t_i = i \Delta t$, $i=0, \dots, n_t$, with $T=n_t \Delta t$ being the time span of the experiment. We consider the experiment parameters $T, n_t$ and $\Delta t$ to be fixed. Experiment design based on varying these parameters are discussed in~\cite{qi2025optimalexperimentaldesignparameter}.


\subsection{Overview}

This paper seeks to achieve two related but distinct objectives which are, (I) improving practical identifiability of a single model that contains undetermined parameters, and (II) discriminating between a pair of models with known parameters, by optimally designing a control input. For each of these objectives, we state the mathematical formulation of the problem, then demonstrate the experimental design procedure using simple ODE models.

In Sec.~\ref{sec:results_iden}, we review the profile likelihood-based approach for identifiability analysis and quantification of parameter uncertainty, and the construction of confidence intervals for parameter values using profile likelihoods. We then demonstrate optimally designing a control input to an experiment to improve the practical identifiability of a model, using the logistic and Richards growth laws for population growth as examples. The main contribution of this section is to demonstrate how to use a profile likelihood-based metric for practical identifiability in the context of optimal experimental design, and how to deal with the associated computational challenges. In Sec.~\ref{sec:results_discrim}, we formulate the model discrimination problem as a open-loop optimal control problem. The control problem, which has an unusual form, is then solved by applying the Pontryagin's Maximum Principle (PMP). We use the same population growth models to illustrate the process of optimal experimental design. This section introduces a new way to formulate the model discrimination problem, and implements an efficient algorithm to solve the optimisation problem. The main novelty of this section, compared to earlier works, is to allow the control function to take a general form, providing greater flexibility in experiment design. Finally, the results are summarised, and their significance is addressed, in Sec.~\ref{sec:exp_design_discussion}.

\section{Optimal experimental design for parameter identifiability}
\label{sec:results_iden}

In this section, we consider using a control to improve the identifiability of a model. We quantify the practical identifiability of a parameter in a model as the width of the 95\% confidence interval for that parameter, which is found via its univariate profile likelihood. The method of profile likelihoods is one of the main techniques for practical identifiability analysis, and has been used in many studies~\cite{liu2024ParameterIdentifiabilityModel,raue2009StructuralPracticalIdentifiability,simpson2022ParameterIdentifiabilityModel,simpson2020PracticalParameterIdentifiabilitya,villaverde2022AssessmentPredictionUncertainty,wieland2021StructuralPracticalIdentifiability}. Here we introduce the approach briefly.

Consider a general ODE model given by Eq.~\eqref{eqn:gen_ode}, and an observation model, Eq.~\eqref{eqn:gen_observation}, 
\begin{align}
    \dv{\bx}{t} &= f(\bx, t, \bu; \btheta), \quad \bx(0) = \bx_0, \label{eqn:gen_ode} \\
    \by_i(\btheta, \bu) &= g(\bx(t_i; \btheta, \bu); \btheta) + \bxi_i, \quad i=0, \dots, n_t, \quad \bxi_i \sim \mathcal{N}(0, \sigma^2), \label{eqn:gen_observation}
\end{align}
where $\bx$ denotes the state variables, $\bu$ the control variables, $\btheta$ the parameters, $\by$ the observables, which are measured at time-points $t_i, i=0, \dots, n_t$, and $\bxi$ the observational noise. Let $\by_{\textrm{model},i}(\btheta,\bu)$ denote the output of the model, and $\by_{\textrm{data},i}$ denote the experimental data. We use the notation $\btheta_{-i}$ to denote the parameter vector with $\theta_i$, the $i\tth$ parameter, removed.
Let $L(\by_{\textrm{data}}|\btheta,\sigma)$ denote the likelihood function, and $\btheta^*, \sigma^*$ be the MLE generated using available data. The normalised univariate profile likelihood function $l_i(\theta_i')$ for parameter $\theta_i$ is defined to be
\begin{equation}
l_i(\theta_i') = \max_{\btheta_{-i}, \sigma} \left[ \log L(\by_{\textrm{data}}| \btheta,\sigma)|_{\theta_i=\theta_i'} \right] - \log L(\by_{\textrm{data}} | \btheta^*, \sigma^*),
\label{eqn:log-profile-likelihood}
\end{equation}
which will henceforth be referred to as the profile likelihood function for brevity. The profile likelihood function can be used to construct an approximate 95\% confidence region (which is usually an interval, but can possibly be a union of disjoint intervals if the profile likelihood is multi-modal) for the value of $\theta_i$~\cite{murphy2000ProfileLikelihood,simpson2022ParameterIdentifiabilityModel},
\begin{equation}
\{\theta_{i} | l_{i}(\theta_i) > -\chi^2(0.95; 1)/2 \approx -1.92\}.
\label{eqn:conf_interval}
\end{equation}
We use $\Delta \theta_i$ to denote the size of the confidence interval, which depends on the control $\bu$. For a given model parameter $\theta_i$ that we would like to estimate, the goal of experimental design is to minimise $\Delta \theta_i$ by optimally choosing a control function, $\bu(t)$.  This problem is rather difficult to formulate as a traditional optimal control problem, as the objective, $\Delta \theta_i$, cannot be easily expressed as a function of the state and control variables. To make the problem easier, we will restrict our attention to the case of a scalar control taking the form of a window function, 
\begin{equation}
u(t) = \left\{
\begin{aligned}
&\umax, \quad &&\tau_0 < t < \tau_0+\tau, \\
&0, &&\text{\ otherwise} ,
\end{aligned}
\right.\label{eqn:control_window_fn}
\end{equation}
where $\umax \geq 0$ is the height of the window function, $0 \leq \tau_0 \leq T$ is the time when the control is turned on, and $\tau \leq T-\tau_0$ is the duration for which the control is turned on. We have effectively parameterised the control with three control parameters, $\umax$, $\tau_0$, and $\tau$. We use MATLAB's \textit{fmincon} optimisation routine to perform the optimisation in the calculations for the profile likelihoods (Eq.~\eqref{eqn:log-profile-likelihood}), and use the \textit{fzero} root-finding routine to find the lower and upper bounds of the confidence intervals, as defined in Eq.~\eqref{eqn:conf_interval}.

\subsection{Logistic growth model}
\label{sec:iden_logistic}

We now illustrate the process for designing an experiment to optimise parameter identifiability with the logistic growth model. Here, we use logistic growth to describe the dynamics of a population of cells in a cell proliferation assay. The model equations are
\begin{align}
    \dv{C}{t} &= rC \left( 1 - \frac{C}{K} \right) - \delta C \coloneqq f_L(C; (r, \delta, K)), \quad C(0)=C_0, 
    \label{eqn:logistic_model_basic} \\
    y_i &= C(t_i) + \xi_i, \quad \xi_i \sim N(0, \sigma^2), \quad i= 1, \dots, n_t,
\end{align}
where $C(t)$ is the number density of cells at time $t$, $y_i$ is the noisy measurement of cell density at time $t_i$, $r>0$ the proliferation rate, $K>0$ the carrying capacity, and $\delta \geq 0$ the rate of cell death. The model parameters are $\btheta=(r, \delta ,K)$. The units for the quantities in the logistic model and the control, used throughout the rest of the paper, are
\begin{equation}
    [C] = [K] = [\sigma] = \textrm{cell/mm}^2, \quad [t] = \textrm{h}, \quad  [r] = [\delta] = \textrm{h}^{-1}, [\tau]=[\tau_0] = \textrm{h}.\label{eqn:logistic_units}
\end{equation}
The logistic model, as usually given in the literature, does not explicitly include the death term, since it can be absorbed into the linear proliferation term. If we define the effective growth rate and carrying capacity as
\begin{equation}
\reff = r - \delta, \quad \keff=K\left(1-\frac{\delta}{r} \right) = K \left(\frac{\reff}{r} \right),
\label{eqn:effective_params}
\end{equation}
then
\begin{equation*}
f_L(C; (r, \delta, K)) = f_L(C; (\reff, 0, \keff)) \ \textrm{for all}\ C\geq 0.
\end{equation*}
There are biologically relevant reasons why we might introduce an explicit death term. For example, it is plausible that the proliferation rate is density-dependent while the death rate is density-independent, or we might be interested in estimating both parameters separately, such as in~\cite{browning2023PredictingRadiotherapyPatient}, where the authors used a linear decay term to represent the necrosis of tumour cells, separate from the logistic growth term. However, the addition of the death term means that the parameters $r, \delta$ and $K$ in both models are structurally non-identifiable, since a change in the value of one of the three can be compensated by a corresponding  change in the value of the other two, by holding $\reff$ and $\keff$ constant.
It can be shown that $\reff$ and $\keff$ are, however, structurally identifiable parameter combinations (see Supplementary Materials~\ref{apx:structural}).

Now, suppose we have the following parameter set as ``ground truth'', representing a best educated ``guess'' of the parameter values before the experiment:
\begin{equation}
\hat{r}=0.45 \text{\ h}^{-1}, \quad \hat{\delta}=0.15 \text{\ h}^{-1}, \quad \hat{\gamma}=1, \quad \hat{K}=3900 \text{\ cell/mm}^{2}, \quad \sigma = 20 \text{\ cell/mm}^{2}.
\label{eqn:default_params_control_iden_logistic}
\end{equation}
Such a ``guess" can be informed by existing knowledge of biological systems similar to the one being studied. These values are chosen so that the effective parameter values, $\reff$ and $\keff$ as defined in Eq.~\eqref{eqn:effective_params}, approximate the parameter values inferred using an experimental dataset from a proliferation assay with MDCK cells in~\cite{liu2024ParameterIdentifiabilityModel}.
The other experimental parameters are set to be
\begin{equation*}
\Delta t = 0.25 \text{\ h}, \quad n_t=100, \quad T = 25 \text{\ h}, \quad C_0=100 \text{\ cell/mm}^{2},
\end{equation*}
which are realistic for such an experiment. We now consider three possible ways to introduce a control into the experiment.

\subsection{Using $u_K$ as the control variable}
\label{sec:exp_design_control_identifiability_uk}

The first way to introduce a control is to assume that we can additively modulate the carrying capacity, $K$. This can be done by, for example, changing the amount of nutrients provided to the cell culture. The model equations (Eq.~\eqref{eqn:logistic_model_basic}) can be modified to
\begin{align}
\dv{C}{t} &=rC \left[ 1 - \frac{C}{K-u_K(t)} \right] - \delta C,
\label{eqn:richards_ode_control_additive_uk}
\end{align}
where $0 \leq u_K(t) < K$ has units cell/mm$^2$, and takes the form of a window function, as in Eq.~\eqref{eqn:control_window_fn}. Applying this control allows us to effectively observe the behaviour of the system in two different conditions: when $u_K=0$, and when $u_K=\umax$. If another parameter set $\btheta=(r, \delta, K)$ yields the same model output as $\Hat{\btheta} = (\hat r, \hat \delta, \hat K)$ in the absence of observational error, then the parameters must satisfy the following three equations:
\begin{subequations}
\label{eqn:practical_noniden_conditions}
\begin{align}
&\reff=r-\delta=\hat{r}-\hat{\delta}, \label{eqn:practical_noniden_conditions1}\\
&\keff=K(1-\delta/r)=\hat{K}(1-\hat{\delta}/\hat{r}), \label{eqn:practical_noniden_conditions2}\\
&(K-\umax)(1-\delta/r)=(\hat{K}-\umax)(1-\hat{\delta}/\hat{r}).\label{eqn:practical_noniden_conditions3}
\end{align}
\end{subequations}
It can be shown that this set of equations imply $\btheta = \Hat{\btheta}$, so in the presence of the control, the model is at least structurally identifiable.

\begin{figure}[H]
    \centering
    \includegraphics[width=\linewidth]{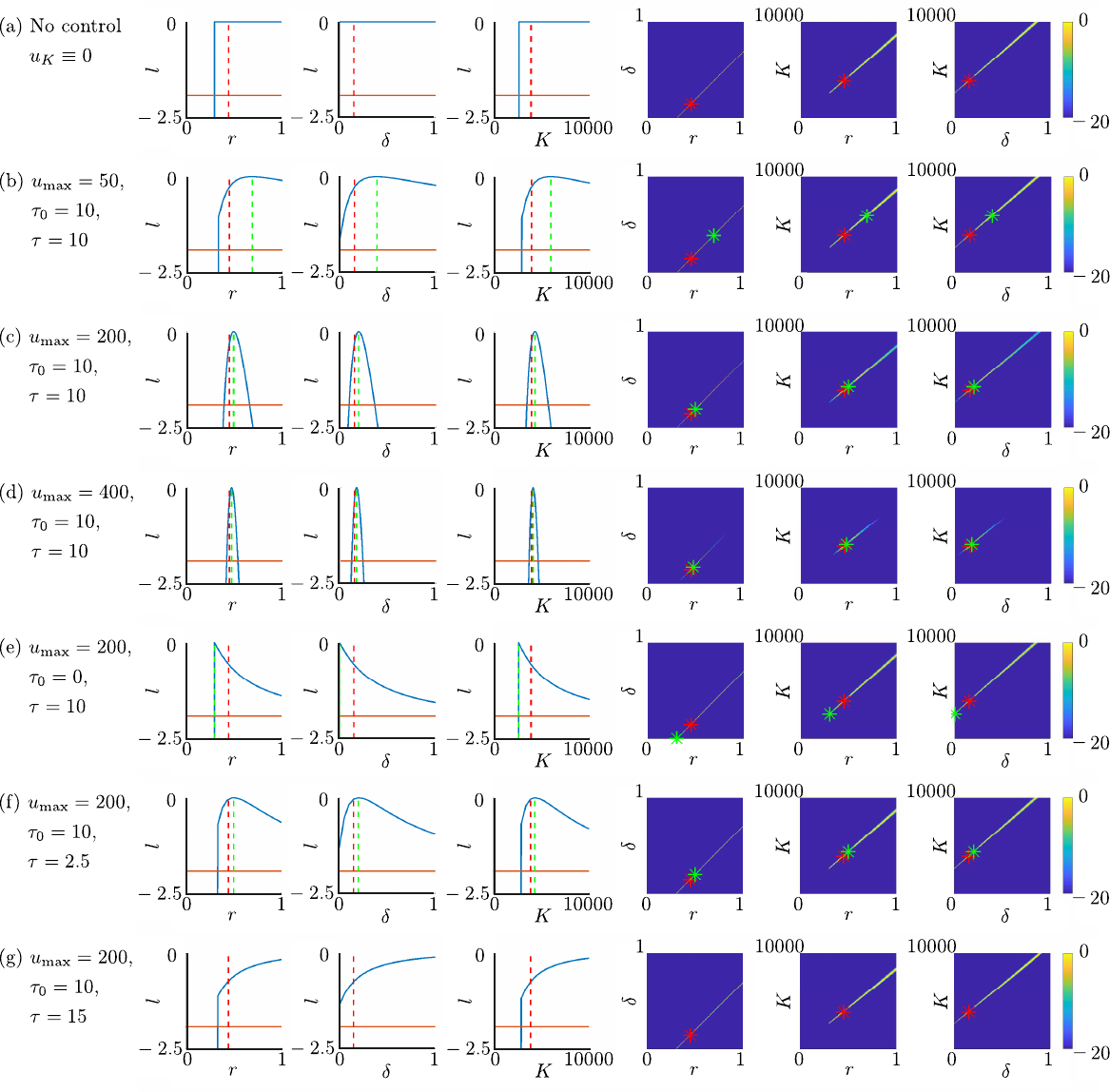}
    \caption[Profile likelihoods for logistic model with additive control with varying control]{The profile likelihood functions for the parameters in the logistic model with $u_K$ as an additive control (Eq.~\eqref{eqn:richards_ode_control_additive_uk}), using a synthetic dataset generated with the parameter values in Eq.~\eqref{eqn:default_params_control_iden_logistic}, with $u_K(t)$ being a window function as in Eq.~\eqref{eqn:control_window_fn} with height $\umax$, turning on at $t=\tau_0$, with a width of $\tau$, and $u_\delta$ and $u_r$ set to zero. Starting from the left, the first three columns show the univariate profile likelihoods for $r$, $\delta$ and $K$, respectively, and the next three columns show the bivariate profile likelihoods for $r$--$\delta$, $r$--$K$, and $\delta$--$K$, respectively. The green dashed lines in the univariate likelihood plots denote the MLEs, and the red dashed lines denote the parameter values used to generate the data. These are denoted with green and red stars in the bivariate plots. The parameter units are as in Eq.~\eqref{eqn:logistic_units}.}
\label{fig:exp_design_control_pl_logistic_uk}
\end{figure}

In Fig.~\ref{fig:exp_design_control_pl_logistic_uk}, we demonstrate how the profile likelihood curves change as the control parameters ($\umax$, $\tau_0$, and $\tau$) are varied. First, in Fig.~\ref{fig:exp_design_control_pl_logistic_uk}(a), we show the profile likelihoods with no control applied. Since the model is known to be structurally non-identifiable, so that a change in one parameter can be perfectly compensated for by changes in other parameters, we obtain a flat top for all three profile likelihood functions, as expected. Nonetheless, we can identify lower bounds for $r$ and $K$. This is because we restrict all parameters to be non-negative, based on the biological interpretations of the model. A value of $r$ below $\reff$, or a value of $K$ below $\keff$, would require a negative $\delta$ to compensate, which is not allowed, hence the profile likelihood functions for $r$ and $K$ drop sharply at these values.

Fig.~\ref{fig:exp_design_control_pl_logistic_uk}(b) shows the results of applying a very weak control, of magnitude $\umax=50$ cells/mm$^2$ (compared to $\hat K=3900$ cells/mm$^2$), in a modestly sized time window. This results in only a marginal improvement in parameter identifiability, as the flat top in the profile likelihoods in Fig.~\ref{fig:exp_design_control_pl_logistic_uk}(a) changes to a curved one, yet the parameters remain practically non-identifiable (recall that we consider a parameter to be practically identifiable  if and only if the confidence interval is finite, and the profile likelihood has a unique global maximum). Increasing the magnitude of the control to $\umax=200$ cells/mm$^2$ while maintaining the same values for $\tau_0$ and $\tau$ (Fig.~\ref{fig:exp_design_control_pl_logistic_uk}(c)) renders all parameters identifiable, and increasing it to $\umax=400$ cells/mm$^2$ (Fig.~\ref{fig:exp_design_control_pl_logistic_uk}(d)) further improves the accuracy of the parameter estimates. In Fig.~\ref{fig:exp_design_control_pl_logistic_uk}(e), the control is applied at the beginning of the experiment, rather than in the middle. Since $K$ has relatively little impact on the dynamics in the early phase of the experiment, placing the window during which $u_K$ is active early on is less effective for improving parameter identifiability than at a later time, and the parameters remain non-identifiable. Finally, Fig.~\ref{fig:exp_design_control_pl_logistic_uk}(f) shows that non-identifiability will also persist if the control is applied too briefly, since there are too little data within the control window to observe the dynamics of the system. On the other hand, applying the control for too long (Fig.~\ref{fig:exp_design_control_pl_logistic_uk}(g)) also results in non-identifiability, but for the opposite reason. In this case, we have too little data outside the window.

Furthermore, we observe that the shapes of the profile likelihood curves of the three parameters tend to be similar, and the widths of the corresponding confidence intervals are highly correlated, as can be seen in the bivariate likelihood plots. In each of those plots, the confidence region, represented by the yellow pixels, forms a very narrow ridge, indicating that the combination of any two parameters is identifiable, even if individually they are not. This is a consequence of the fact that $\reff$ and $\keff$ are both structurally identifiable parameter combinations. For this reason, it is sufficient to look at the identifiability of $r$ to understand the overall identifiability of the whole model. Therefore, the rest of this section will focus on $r$.

\begin{figure}[H]
    \centering
    \includegraphics[width=0.7\textwidth]{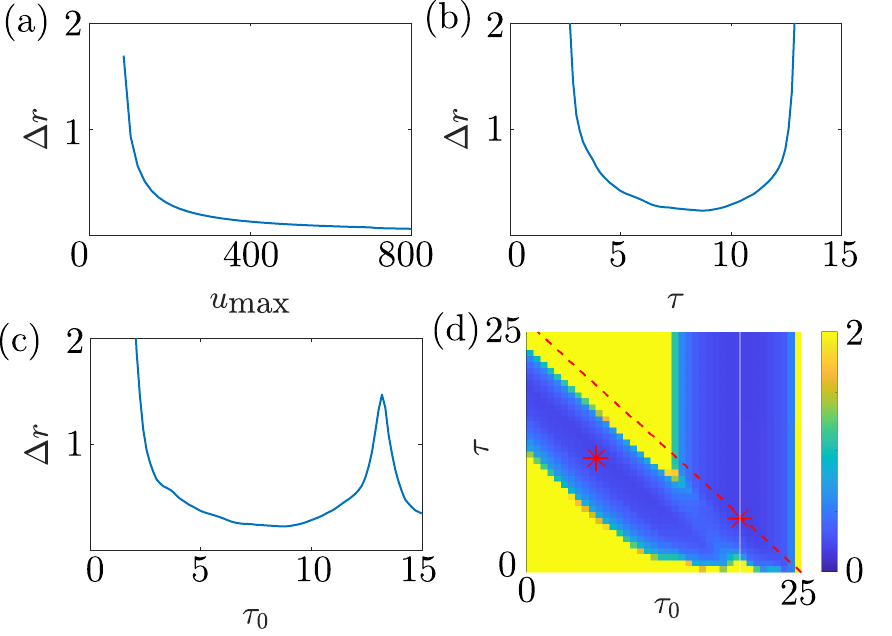}
    \caption[Widths of confidence intervals for the logistic model with additive control $u_K$ as univariate function of control parameters]{The widths of the confidence intervals for $r$, $\Delta r$, in the logistic model with additive control $u_K$ (Eq.~\eqref{eqn:richards_ode_control_additive_uk}), using the synthetic dataset generated with the parameter values in Eq.~\eqref{eqn:default_params_control_iden_logistic}, as a function of the control parameters for $u_K$ defined in Eq.~\eqref{eqn:control_window_fn}. (a-c) $\Delta r$ as a function of a single varying control parameter. The control parameter being varied is labelled on the $x$-axis in each plot, while the other control parameters are held fixed at $\umax=200$ cells/mm$^2$, $\tau_0=10$ h, $\tau=10$ h. Units are in Eq.~\eqref{eqn:logistic_units}. (d) $\Delta r$ shown as a function of $\tau_0$ and $\tau$, represented as a heat-map, with $\umax=1200$ cells/mm$^2$ fixed. The red dashed line marks the line $\tau_0+\tau=25$ h. The two red stars denote local minima of $\Delta r$ at $(\tau_0, \tau)=$ (6.3 h, 12.4 h), (19.5 h, 5.5 h).}
    \label{fig:exp_design_control_uk_rrange_1}
\end{figure}

Next, we explore the effects of the control on identifiability more systematically by plotting $\Delta r$, the width of the confidence interval of $r$, as a function of the control variables. First, in Fig.~\ref{fig:exp_design_control_uk_rrange_1}(a-c), we take $\umax=200$ cells/mm$^2$, $\tau_0=10$ h, $\tau=10$ h which, as shown in Fig.~\ref{fig:exp_design_control_pl_logistic_uk}, makes the model parameters identifiable. Then, we vary one control parameter at a time while leaving the other two fixed, and plot $\Delta r$ as a function of the varying parameter. We observe that $\Delta r$ decreases monotonically as $\umax$ increases, confirming our earlier observations, while the effects of $\tau_0$ and $\tau$ on $\Delta r$ are more nuanced. There seem to be two good choices for $\tau_0$: one around $\tau_0=9$ h and another around $\tau_0=15$ h, which are the local minima of $\Delta r$ as a function of $\tau_0$. We will later provide an intuitive explanation for this using local sensitivity.

In order to further look into the effects of the location of the control window on parameter identifiability, we fix $\umax$ and plot $\Delta r$ as both $\tau_0$ and $\tau$ vary in Fig.~\ref{fig:exp_design_control_uk_rrange_1}(d), which contains Fig.~\ref{fig:exp_design_control_uk_rrange_1}(b,c) as cross-sections. We can ignore the part of the plot above the red diagonal dashed line as it represents the region where $\tau_0+\tau>T$, in which case the control stays on past the end of the experiment, which is no different than the case of $\tau_0 + \tau = T$.
We observe that there are two diagonal valleys (darkest blue) representing the combinations of $(\tau_0, \tau)$ that give the highest degree of identifiability. One of these two valleys represents $\tau_0<20$ h and $\tau_0+\tau \approx 19$ h, and the other represents $\tau_0>13$ h and $\tau_0+\tau \approx 25$ h.


Altogether, the results show that in order to achieve identifiability, the magnitude of the control must be high enough, and the timing and duration of the control window $[\tau_0, \tau_0+\tau]$ must be chosen carefully, so that the span of time where the system behaviour depends strongly on the parameter being acted upon by the control ($K$ in this case), overlaps significantly with both the time span inside the window when the control is turned on, which is $[\tau_0, \tau_0+\tau]$, and the time span outside of it.

We can understand this intuitively by thinking about the source of the non-identifiability.
For any set of parameter values, we need sufficient data on the violation of at least one condition in Eq.~\eqref{eqn:practical_noniden_conditions} in order to confidently conclude that it does not reflect the data. If $\umax$ is not sufficiently high, then any set of parameters that satisfy Eq.~\eqref{eqn:practical_noniden_conditions2} would also approximately satisfy Eq.~\eqref{eqn:practical_noniden_conditions3}, leaving us with effectively only two conditions. If the control window is too short, then we will not have enough data on the violation of Eq.~\eqref{eqn:practical_noniden_conditions3}. Similarly, if the control window is too large, leaving us little data on system behaviour outside of the window, then we would not be able to conclude a violation of Eq.~\eqref{eqn:practical_noniden_conditions2}. In all of these cases, we have two conditions and three unknown parameters, preventing unique identification of parameter values.

\begin{figure}[htb!]
    \centering
    \includegraphics[width=0.85\textwidth]{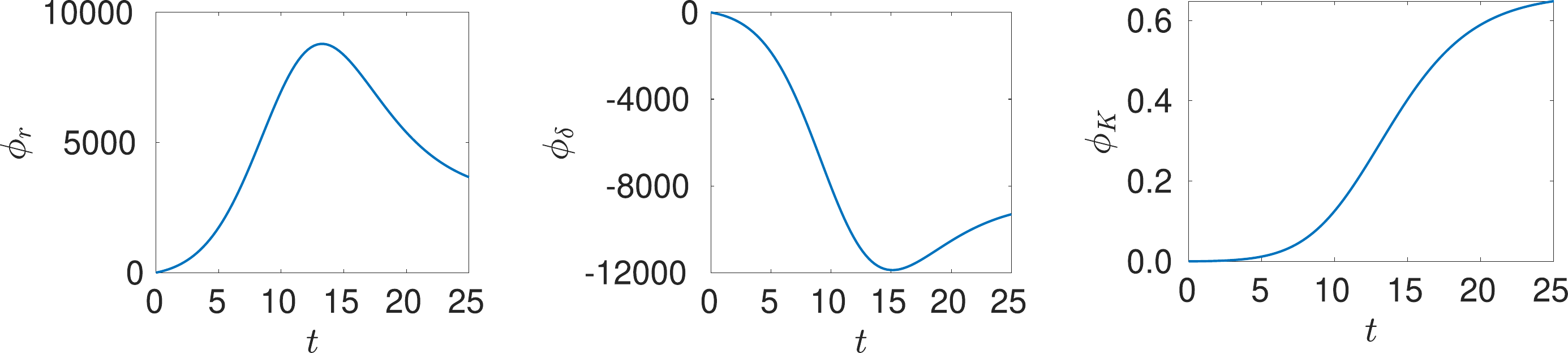}
    \caption[Local sensitivities of logistic model]{Local sensitivities of the logistic model (Eq.~\eqref{eqn:logistic_model_basic}) with parameter values from Eq.~\eqref{eqn:default_params_control_iden_logistic}, with respect to the model parameters $r,\ \delta, \ K$, as defined in Eq.~\eqref{eqn:local_sensitivity}. Units are $[t]=\textrm{h}, [\phi_r]=[\phi_\delta]=\textrm{cells}\cdot\textrm{h}/\textrm{mm}^2, [\phi_K]=1$.}
    \label{fig:sensitivity}
\end{figure}

This idea provides a heuristic approach to approximate the optimal controls in terms of parameter sensitivity. Let $\phi_{\theta_j}(t)$ be the local sensitivity of the solution with respect to parameter $\theta_j$, defined as
\begin{equation}
\phi_{\theta_j}(t) = \pd{\Cmodel(t; \btheta, C_0)}{\theta_j},
    \label{eqn:local_sensitivity}
\end{equation}
where $\Cmodel$ is the output of the model, Eq.~\eqref{eqn:logistic_model_basic}. The sensitivities are calculated using the following ODE, which can be derived from  Eq.~\eqref{eqn:logistic_model_basic} using the chain rule:
\[\dv{\phi_{\theta_j}}{t} = \pd{f(C; \btheta)}{C}\phi_{\theta_j} + \pd{f(C; \btheta)}{\theta_j} , \quad
\phi_{\theta_j}(0) = 0.\]
We plot the sensitivities with respect to $r, \delta$ and $K$ in Fig.~\ref{fig:sensitivity}. A larger value of $|\phi_\theta(t)|$ means the solution is more sensitive to a change in the parameter $\theta$ at time $t$, while holding the other parameter values constant. We have $\phi_\theta(0)=0$, since the initial condition is fixed. For the sake of this heuristic, let us define the sensitive interval for parameter $\theta$ as
\begin{equation}
\mathcal{T}_{\theta_j} = \left\{t \in [0,T]: |\phi_{\theta_j}(t)| \geq \frac{1}{2} \max_{t' \in [0,T] } |\phi_{\theta_j}(t')| \right\}.
\label{eqn:sensitive_interval}
\end{equation}
This interval represents the time span during which the solution is relatively more sensitive to the parameter.
For $K$, we have $\mathcal{T}_K \approx$ [13 h, 25 h]. As a heuristic for optimal control, we should ensure that the control window $[\tau_0, \tau_0+\tau]$ covers roughly half of $\mathcal{T}_K$, so that we have sufficient time to observe the system both when the control is on, and when it is off. This condition gives us two sensible options: one with $\tau_0<13.5$ h and $\tau_0+\tau \approx 19$ h, and another with $\tau_0 \approx 19$ and $\tau_0+\tau=25$. These two options each correspond to a local minimum of $\Delta r$ with respect to $\tau_0$ and $\tau$ as we found in Fig.~\ref{fig:exp_design_control_uk_rrange_1}(d).

Since the local sensitivities $\phi_\theta$, and therefore $\mathcal{T}_\theta$, are much cheaper to compute compared to $\Delta \theta$, this heuristic provides a cheaper alternative to the profile likelihood-based approach. A major disadvantage of it, however, is that it only uses local information in the parameter space, which is the same major weakness of using the FIM to perform identifiability analysis, hence all criticisms of FIM also apply here. Furthermore, the heuristic method does not take into account the ways the parameters compensate for each other (which is encoded in the second mixed partial derivatives of the likelihood function with respect to the parameters, that is, the off-diagonal elements of the FIM). Furthermore, local sensitivity analysis by itself can be misleading---this is illustrated in Sec.~\ref{sec:exp_design_control_identifiability_ud}.

\subsection{Using $u_\delta$ as the control variable}
\label{sec:exp_design_control_identifiability_ud}

We now consider applying a different control, $u_\delta$, to the experiment, where we additively modulate the death rate, $\delta$. A plausible way to apply this control is to add a toxin that kills cells. The model equations (Eq.~\eqref{eqn:logistic_model_basic}) in the presence of $u_\delta$ can be written as
\begin{align}
\dv{C}{t}=rC \left[ 1 - \frac{C}{K} \right] - (\delta+u_\delta(t)) C,
\label{eqn:richards_ode_control_additive_ud}
\end{align}
where $u_\delta$ has units h$^{-1}$. Applying the same reasoning from the beginning of Sec.~\ref{sec:exp_design_control_identifiability_uk}, it can be shown that any parameter set $\btheta$ can produce the same model output as $\Hat{\btheta}$ if the following two equations are satisfied:
\begin{align*}
&\reff=r-\delta=\hat{r}-\hat{\delta},\\
&\keff=K(1-\delta/r)=\hat{K}(1-\hat{\delta}/\hat{r}).
\end{align*}
These two equations are not sufficient to uniquely determine $(r, \delta, K)$, and therefore the model remains structurally non-identifiable in the presence of $u_\delta$. This means that applying $u_\delta$ as an additive control will not improve parameter identifiability.
However, this is not revealed by local sensitivity analysis (Fig.~\ref{fig:sensitivity}(b)), which suggests that applying $u_\delta$ in the early stages of the experiment might be useful for improving parameter identifiability. This demonstrates that local sensitivity analysis, while a useful heuristic, cannot fully replace identifiability analysis for experimental design.

\subsection{Using $u_r$ as the control variable}
\label{sec:exp_design_control_identifiability_ur}

We consider one last control input to the experiment, $u_r$, which additively modulates the proliferation rate of the cells, $r$. This can be done by e.g. applying a certain signalling protein that accelerates the cell cycle, therefore speeding up growth, corresponding to a temporarily elevated $r$. The model equations (Eq.~\eqref{eqn:logistic_model_basic}) in the presence of $u_\delta$ can be written as
\begin{align}
\dv{C}{t} &=(r+u_r(t))C \left[ 1 - \frac{C}{K} \right] - \delta C,
\label{eqn:richards_ode_control_additive_ur}
\end{align}
where $u_r(t)$ has units h$^{-1}$. With $u_r$ as the control variable, the conditions for two parameter sets $\btheta =(r,\delta, K)$ and $\Hat{\btheta}=(\hat r, \hat \delta, \hat K)$ to produce the same model output in the absence of observational noise is
\begin{subequations}
\label{eqn:practical_noniden_conditions4_all}
\begin{align}
&\reff=r-\delta=\hat{r}-\hat{\delta}, \\
&\keff=K(1-\delta/r)=\hat{K}(1-\hat{\delta}/\hat{r}), \\
&K\left(1- \frac{\delta}{r+\umax} \right) = \hat{K}\left(1- \frac{\hat{\delta}}{\hat{r}+\umax} \right),\label{eqn:practical_noniden_conditions4}
\end{align}
\end{subequations}
which are sufficient to ensure that $\btheta = \Hat{\btheta}$, and therefore the addition of $u_r$ makes the model parameters structurally identifiable.

\begin{figure}[H]
    \centering
    \includegraphics[width=0.7\textwidth]{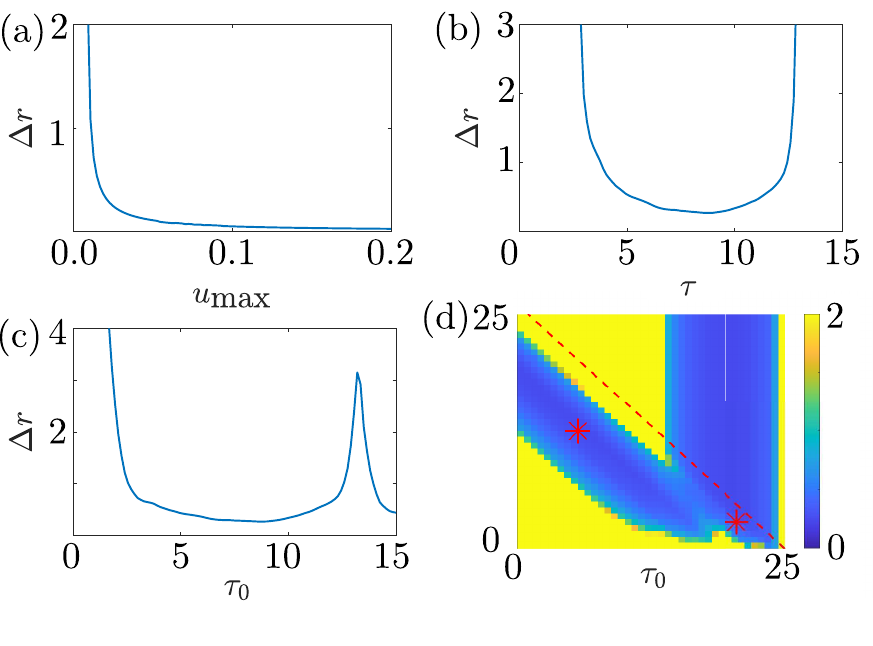}
    \caption[Widths of confidence intervals for the logistic model with additive control $u_r$ as function of single control parameters]{Similar to Fig.~\ref{fig:exp_design_control_uk_rrange_1}, except the applied control variable is $u_r$ instead of $u_K$. 
    (a-c) The widths of the confidence intervals of $r$, $\Delta r$ are shown as a function of one of the control parameters for $u_r$ as defined in Eq.~\eqref{eqn:control_window_fn}, with the other two held fixed at $\umax=0.02, \tau_0=10, \tau=10$. The units are given in Eq.~\eqref{eqn:logistic_units}. (d) $\Delta r$ shown as a function of $\tau_0$ and $\tau$, represented as a heat-map, with $\umax=0.02$ fixed. The red dashed line marks the line $\tau_0+\tau=25$. The two red stars denote the local minima of $\Delta r$ at $(\tau_0, \tau)=$(5.6 h, 13.0 h), (20.5 h, 4.5 h).
    }
    \label{fig:exp_design_control_rrange_ur_1}
\end{figure}

As in Sec.~\ref{sec:exp_design_control_identifiability_uk}, we explore how $\Delta r$ changes as the control parameters change. We centre our explorations around $\umax=0.02$ h$^{-1}$, $\tau=10$ h, $\tau_0=10$ h, which makes the model identifiable, and vary one or two of these parameters at a time. 
The results are plotted in Fig.~\ref{fig:exp_design_control_rrange_ur_1}.

We observe that the overall shape of $\Delta r$ in the $u_r$ case shown in Fig.~\ref{fig:exp_design_control_rrange_ur_1} is similar to that for $u_K$, shown in Fig.~\ref{fig:exp_design_control_uk_rrange_1}.  A higher $\umax$, the magnitude of the control, is always helpful for improving identifiability. The combinations of $(\tau_0,\tau)$ that give the highest degree of identifiability satisfy either $\tau_0<16$ h and $\tau_0+\tau \approx 18$ h, or $\tau_0>16$ h and $\tau_0+\tau \approx 25$ h. The sensitive interval (Eq.~\eqref{eqn:sensitive_interval}) for $r$ is $\mathcal{T}_r \approx [8,22]$, and the above combinations represent windows that roughly overlap with half of this interval, so our sensitivity-based heuristic is still valid in this case.

The results in this section show that an intelligent choice of the control variable, along with a strategic choice of the control, can be effective at removing structural non-identifiability and improving practical identifiability. This choice can be informed by using profile likelihoods to inspect the dependence of uncertainty in parameter estimates on the control. We have also shown a heuristic way to select the control, which involves computing local sensitivities of the parameters and some simple calculations. This heuristic is intuitive to understand, and is much less computationally expensive, but should be used only in conjunction with structural identifiability analysis.

\section{Optimal experimental design for model discrimination}\label{sec:results_discrim}

In this section, we seek to optimally distinguish two models that do not contain undetermined parameters, i.e. to enable model selection, by designing a control input.  Suppose that the two models can be written as
\begin{equation}
    \dv{\bx_i}{t} = f_i(\bx_i,t,\bu), \quad \by_{i}=g_i(\bx)+\bxi_{i}, \quad \bx_i(0)=\bx_0, \quad i=1,2,
    \label{eqn:control_state_eqn}
\end{equation}
and $\bx_{i}(t; \bu(t))$ is the solution to these models given the control input $\bu(t)$. Since the two models describe the same system, we assume that they share the same initial condition. We wish to maximise the squared difference between model measurements over time, that is, we want to find $\bu(t)$ to maximise the cost functional
\begin{align}
    \argmax_{\bu \in \mathcal{U}} J[\bu]&= \argmin_{\bu \in \mathcal{U}}\int_0^T  \Lag(\bx,\bu) \text{d}t,\notag \\
    \Lag(\bx,\bu) &= -\left[g_1(\bx_{1}(t; \bu(t))) -g_2(\bx_{2}(t; \bu(t)))\right]^2 + \balpha \cdot \bu(t),
    \label{eqn:optimal_control_discrimination_problem}
\end{align}
where $\mathcal{U}$ denotes the set of values the control variable is allowed to take, and $\balpha$ encodes the cost of applying the control. We note that, in practice, the measurements are noisy, and made at discrete time points (e.g. $\by_{i,j} = g(\bx_i(t_j))$), and the integrals should be replaced with summation. However, for simplicity, we assume that we have access to noiseless measurements in continuous time. The integrand $\Lag$ is known as the Lagrangian, it is assumed to be continuous in all its inputs, and all of its partial derivatives are assumed to be continuous. We consider both the case of continuous control, and bang-bang control. In continuous control, each of the control variables may take any value between zero and an imposed upper bound. The upper bounds represent what is realistically achievable in an experimental setting, or a ``safe" level beyond which we might alter the system dynamics too much for the model to be valid.
In bang-bang control, the control variable is a piecewise constant function that may only take the value of zero or the upper bound. While continuous control allows more flexibility and greater room for optimisation, in an experiment it is often difficult to fine-tune the precise value of the control, so bang-bang control may be more applicable. For this reason, bang-bang control has seen wide-ranging applications, such as cancer chemotherapy~\cite{ledzewicz2002OptimalBangBangControls} and viral treatment~\cite{mapder2019PopulationBangbangSwitches}. 

To solve this optimal control problem, we apply Pontryagin's Maximum Principle (PMP), using the Forward-Backward Sweep (FBS) algorithm to solve the resulting systems of equations numerically~\cite{lenhart2007OptimalControlApplied,liberzon2011CalculusVariationsOptimal,pmp,sharpOptimalControlAcute2019}. 
The details of the FBS algorithm are given in Supplementary Materials~\ref{apx:pmp}.
We will now illustrate the aforementioned approach for experimental design by applying PMP with the FBS algorithm to optimally discriminate between models of cell proliferation.

\subsection{Logistic model with additive controls}\label{sec:discrim_logistic_additive}

First, consider discriminating between two instances of the logistic model with the following two parameter sets:
\begin{subequations}
\label{eqn:default_params_control_discrim_logistic}
\begin{align}
    \dv{C_1}{t} &= 0.45 C_1 \left(1-\frac{C_1}{3900} \right) - 0.15 C_1 = f_L(0.45, 0.15, 3900), \\
    \dv{C_2}{t} &= 0.3 C_2 \left(1-\frac{C_2}{2600} \right) = f_L(0.3, 0, 2600), 
\end{align}
\end{subequations}
where $f_L$ is from Eq.~\eqref{eqn:logistic_model_basic}. The timespan of the experiment is taken to be $T=25$ h, and the initial condition is $C(0)=100$ cells/mm$^2$. The first parameter set was chosen to coincide with those in Eq.~\eqref{eqn:default_params_control_iden_logistic}, and the second parameter set has the same effective parameter values $\reff$ and $\keff$ (Eq.~\eqref{eqn:effective_params}) as the first parameter set, which means they give rise to identical solutions without control. We might be interested in discriminating these two models from a biological perspective to determine whether the mechanism represented by the linear death term genuinely exists within the system.

We first consider the model with additive controls, as we have done in Sec.~\ref{sec:results_iden}.
As discussed in Sec.~\ref{sec:exp_design_control_identifiability_ud}, applying $u_\delta$ additively does not remove structural identifiability, as both models yield the exact same solution for any $u_\delta$. Therefore, we consider only applying $u_K$ or $u_r$. 

When applying $u_K$ additively, the model equations become
\begin{subequations}
\label{eqn:logistic_discrimination_additive}
\begin{align}
    \dv{C_1}{t} &= 0.45 C_1 \left(1-\frac{C_1}{3900-u_K(t)} \right) - 0.15 C_1, \\
    \dv{C_2}{t} &= 0.3 C_2 \left(1-\frac{C_2}{2600-u_K(t)} \right).
\end{align}
\end{subequations}
We take $u_{K,\textrm{max}}=1200$ cells/mm$^2$ and $\alpha_K=0.03$ cells/mm$^2$. A much larger $\alpha_K$ (e.g.,~$\alpha_K=0.1$ cells/mm$^2$) results in $u_K^*(t) \equiv 0$, as the cost of the control is too great to be compensated by the reward of being able to discriminate the models. On the other hand, a much smaller $\alpha_K$ (e.g.,~$\alpha_K=0.005$ cells/mm$^2$) results in $u_K^*(t) \equiv u_{K,\textrm{max}}$, as a larger $u_K$ is always helpful in discriminating between the two models in this case, and the cost of turning $u_K$ on is negligible. Therefore, there is a range for $\alpha_K$ to make the problem non-trivial. 

\begin{figure}[hbt!]
     \centering
     \includegraphics[width=0.9\textwidth]{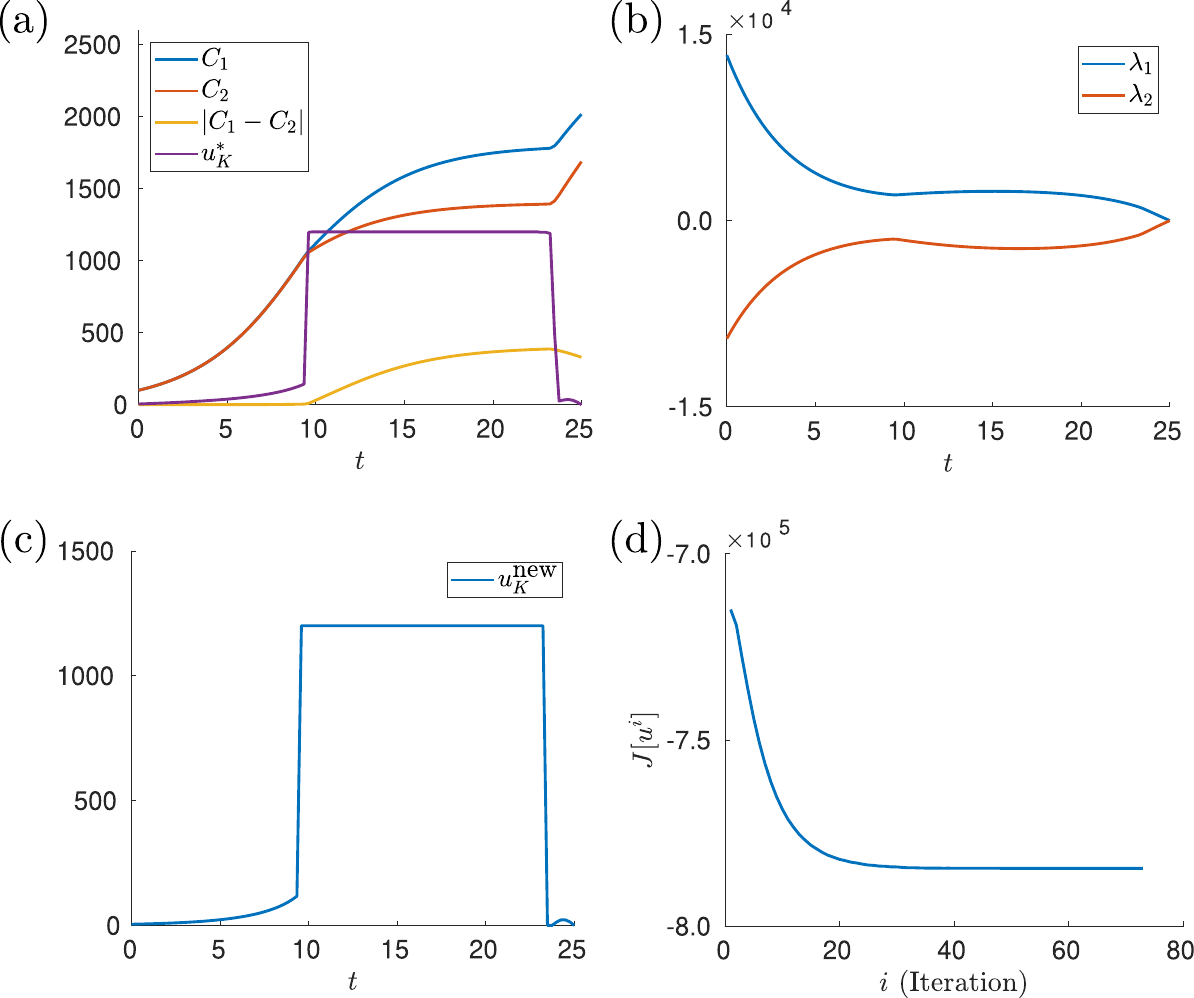}
        \caption[Optimal control for model discrimination, with additive $u_K$ as control variable]{Results for discriminating between logistic models with additive $u_K$ as control (Eqs.~\eqref{eqn:optimal_control_discrimination_problem},~\eqref{eqn:logistic_discrimination_additive}), $u_{K,\textrm{max}}=1200$ cells/mm$^2$, and $\alpha_K=0.03$ cells/mm$^2$. (a) The optimal control $u_K^*(t)$, as found by the algorithm, and the corresponding trajectories of the state variables $C_1, C_2$, as well as their difference over time. (b) The trajectories of the adjoint variables $\lambda_1, \lambda_2$ corresponding to the optimal control. (c) $u_K^{\textrm{new}}$, the update to the control, computed at the final iteration before termination of the algorithm. It is very close to $u_K^*$, suggesting that further iterations will not lead to significant changes in the control. (d) The cost functional $J[u]$ over the iterations, showing its convergence near the final iterations. The units are given in Eq.~\eqref{eqn:default_params_control_iden_logistic}, and $[u_K]=$cells/mm$^2, [J]=(\textrm{cells/mm}^2)^2$.}
        \label{fig:exp_design_control_discrim_logistic_uk}
\end{figure}

In Fig.~\ref{fig:exp_design_control_discrim_logistic_uk}, we show the result of applying the FBS algorithm to the optimal control problem, with $u_K$ the control variable. The strategy implied by the resulting optimal control can be summarised as allowing both solutions to reach roughly $\keff/2$ without much intervention, then abruptly turning on the control. In the early stages of the experiment (for small $t$), the carrying capacity $K$ (and therefore the control variable $u_K$) has relatively little impact on the solutions of both models, so the control remains low. From $t \approx 9.5$ h, the impact of the control on discriminating the two model solutions is large enough to overcome the cost of applying the control, and $u_K$ is turned on to the maximum, where it remains until $t\approx 23.3$ h. After this point, the control  is almost turned off completely until $t=T$, when it is exactly zero, as we know it must be. We can relate this control strategy to the local sensitivity, $\phi_K$. In Fig.~\ref{fig:sensitivity}(c), we can observe that $\phi_K$ is monotonically increasing with time. The period of time where $u_K^*=u_{K,\textrm{max}}$ roughly coincides with the time window where $\phi_K$ is large. This is expected, since the control is likely to be most effective in distinguishing between the model solutions at points where the models are sensitive to the parameter associated with the control. Interestingly, the optimal control has a small ``bump'' near $t\approx 24$ h. The reason for its presence is not immediately apparent, however removing the ``bump'' results in a higher cost functional, suggesting that it is not simply a numerical artifact.

\begin{figure}[H]
    \centering
    \includegraphics[width=0.7\textwidth]{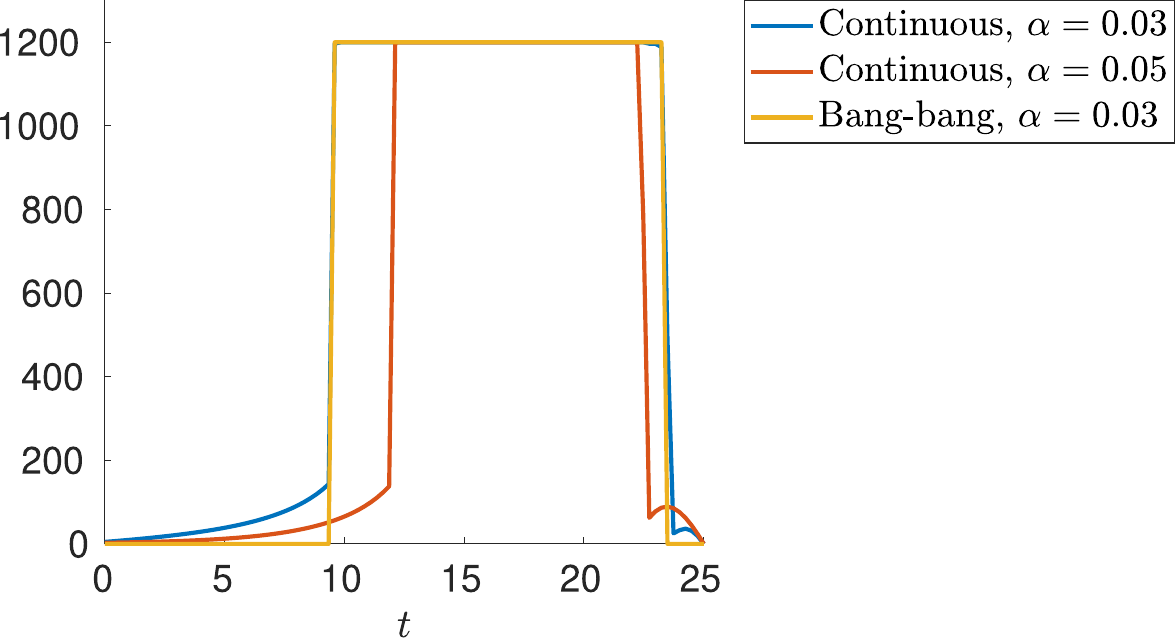}
    \caption[Comparison of continuous and bang-bang optimal controls for model discrimination]{Comparison of the solutions of variations of the optimal control problem. The blue curve is the same control as in Fig.~\ref{fig:exp_design_control_discrim_logistic_uk}. The orange curve is the optimal continuous control with a different value for the weight of control cost, $\alpha$, and the yellow curve is the optimal bang-bang control using the same value of $\alpha$ as the blue curve. The units are $[t]=\textrm{h}, [u_K]=$ cells/mm$^2$.}
    \label{fig:exp_design_control_discrim_logistic_uk_misc}
\end{figure}

Next, we consider a few variations of the optimal control problem, to determine how various factors impact the optimal control.
Changing the value of $\alpha_K$ does not result in a qualitative change in the optimal control. Fig.~\ref{fig:exp_design_control_discrim_logistic_uk_misc} shows that with a higher value of $\alpha_k=0.05$ cells/mm$^2$, the control is turned on over a shorter window between $t\approx 12$ h and $t\approx 22.3$ h, but retains the same shape as in Fig.~\ref{fig:exp_design_control_discrim_logistic_uk}. The small bump near $t=T$ is still present.
The optimal bang-bang control for the same $\alpha$, as found by direct optimisation, is turned on at $t\approx 9.4$ h, almost the same time as when the optimal continuous control is ramped up sharply to its maximum, and turned off at $t \approx 23.4$ h, very close to the time when the optimal continuous control decreases sharply to near zero. The bang-bang optimal control achieves an objective value of $J=-779401$ (cells/mm$^2$)$^2$, only slightly worse than the continuous optimal control, which is $J=-784470$ (cells/mm$^2$)$^2$.
These results show that in this case, we do not gain much from allowing the control to take continuous values, and a bang-bang optimal control is both easier to compute, and experimentally easier to implement.

The optimal control for model discrimination is quite different from the optimal control for improving parameter identifiability, discussed in Sec.~\ref{sec:exp_design_control_identifiability_uk}. This highlights that model discrimination and improving identifiability are distinct problems. In the model discrimination problem, we are essentially proposing a new experiment in addition to the experiment with which we calibrated our two models. For the new experiment to be the most useful, we want to observe the system behaviour in an environment as different to the first experiment as possible, i.e.,~with the control turned to a maximum, whereas in the problem of identifiability improvement, we are essentially conducting two experiments at once, so the data should be split between when the control is on or off.

\begin{figure}[H]
    \centering
    \includegraphics[width=0.7\textwidth]{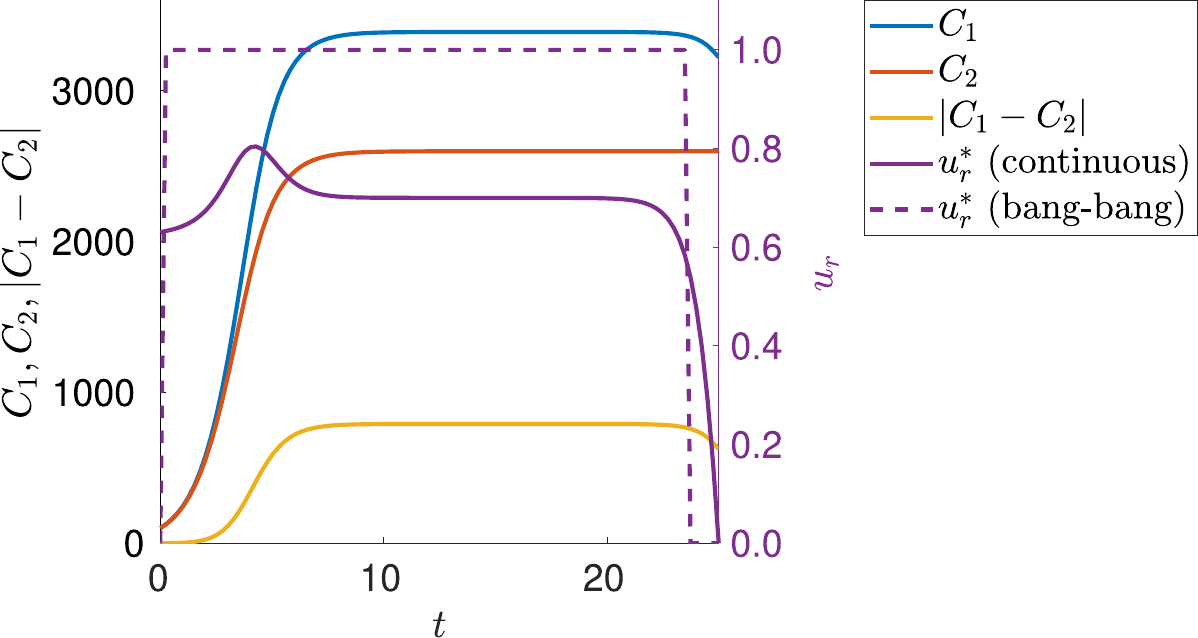}
    \caption[Optimal controls model discrimination, with additive $u_r$ as control variable]{Results of FBS (Algorithm~\ref{alg:fbs}) applied to the optimal control problem for discriminating between logistic models with additive $u_r$ as control (Eqs.~\eqref{eqn:optimal_control_discrimination_problem},~\eqref{eqn:logistic_discrimination_additive_ur}), $u_{r,\textrm{max}}=1$ h$^{-1}$, and $\alpha_r=500000$ h(cells/mm$^2$)$^2$, showing the optimal control $u_K^*(t)$ found by the algorithm, the corresponding trajectories of the state variables $C_1, C_2$, and their difference over time.
    The optimal bang-bang control is also shown for comparison.
    The units are $[C]=\textrm{cells/mm}^2, [t]= \textrm{h}, [u_r]=\textrm{h}^{-1}$.}
    \label{fig:exp_design_control_discrim_logistic_ur}
\end{figure}

Next, we consider the case of applying $u_r$ as the control variable, with the following model equations:
\begin{subequations}
\begin{align}
    \dv{C_1}{t} &= (0.45+u_r(t)) C_1 \left(1-\frac{C_1}{3900} \right) - 0.15 C_1, \\
    \dv{C_2}{t} &= (0.3+u_r(t)) C_2 \left(1-\frac{C_2}{2600} \right).
\end{align}
\label{eqn:logistic_discrimination_additive_ur}
\end{subequations}
The optimal control found by the FBS algorithm is shown in Fig.~\ref{fig:exp_design_control_discrim_logistic_ur}. The optimal control is roughly constant, at $u_r \approx 0.7$ h$^{-1}$ over most of the domain, with a noticeable bump near the inflection point of the solutions, and ramps down smoothly to zero as $t \to T$. 
The strategy seems to be to turn on the control early so that both solutions reach their carrying capacity quickly, then maintain the control in order to observe the difference in $\keff$ for the two models. 

The reason for a local maximum in $u_r^*$ near the inflection point of the two model solutions is not immediately clear, but a possible explanation can be given using local sensitivity. 
Examine the plot of local sensitivity $\phi_r$ (Fig.~\ref{fig:sensitivity}(a)) and the associated model solution (yellow curve in Fig.~\ref{fig:model_discrim}), we can observe that the inflection point of the model solution is near the global maximum of $\phi_r$. Therefore, the local maximum of the control, $u_r^*$, is close to where both model solutions are most sensitive to the parameter $r$, which is sensible for the objective of distinguishing the two model solutions.
The window in time where the optimal bang-bang control is turned on covers most of the time span of the simulation, but it is turned off a short time before $t=T$, around the same time as the optimal continuous control is ramped down. The bang-bang control achieved an objective value of $J=-4618503$ (cells/mm$^2$)$^2$, compared to $J=-6388826$ (cells/mm$^2$)$^2$ for the optimal continuous control (dashed purple line in Fig.~\ref{fig:exp_design_control_discrim_logistic_ur}), which is a noticeably worse performance.

\subsection{Logistic model with multiplicative controls}
\label{sec:discrim_logistic_multiplicative}

Next, we again consider the case for distinguishing between two instances of the logistic model, but now the controls are assumed to affect the associated parameters multiplicatively, rather than additively. While this is a small conceptual difference compared to the previous problem in Sec.~\ref{sec:discrim_logistic_additive}, it nonetheless results in significant changes in the outcome. The model equations (Eq.~\eqref{eqn:default_params_control_discrim_logistic}) in the presence of multiplicative controls can be written as
\begin{subequations}
\label{eqn:logistic_discrimination_multiplicative}
\begin{align}
    \dv{C_1}{t} &= 0.45(1+u_r(t)) C_1 \left(1-\frac{C_1}{3900(1-u_K(t))} \right) - 0.15(1+u_\delta(t)) C_1, \\
    \dv{C_2}{t} &= 0.3(1+u_r(t)) C_2 \left(1-\frac{C_2}{2600(1-u_K(t))} \right).
    \label{eqn:multiplicative_model2}
\end{align}
\end{subequations}
Note that Eq.~\eqref{eqn:multiplicative_model2} does not contain $u_\delta$ because $\delta=0$ in that model, so a multiplicative $u_\delta$ has no effect.
The control variables $u_r, u_\delta, u_K$ are dimensionless in this formulation.
As before, we only consider applying one control variable at a time, while setting the other two to zero. 

\begin{figure}[H]
     \centering
     \includegraphics[width=0.8\textwidth]{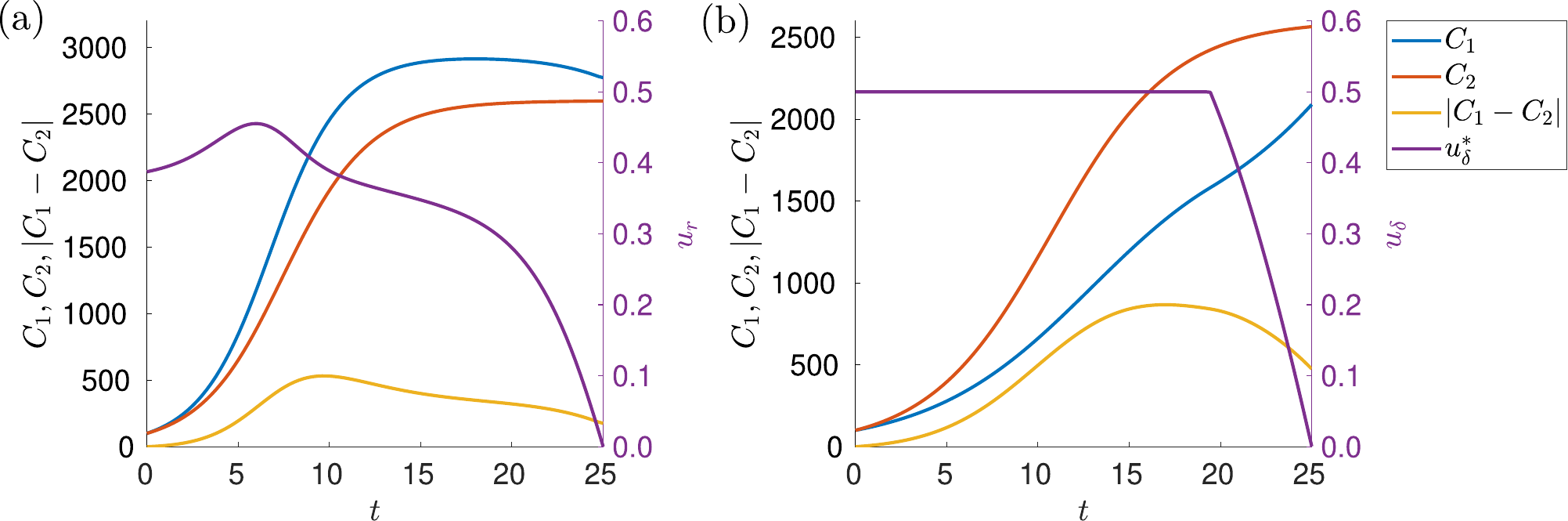}
        \caption[Optimal controls for model discrimination, with multiplicative control variables]{Results of the FBS (Algorithm~\ref{alg:fbs}) applied to the optimal control problem for discriminating between logistic models with multiplicative controls (Eq.~\eqref{eqn:optimal_control_discrimination_problem},~\eqref{eqn:logistic_discrimination_multiplicative}).
        (a) The control variable is $u_r$, with $u_r^{\textrm{max}}=0.5$, $\alpha_r=700000$ (cells/mm$^2$)$^2$. Here $u_\delta = u_K \equiv 0$.
        (b) The control variable is $u_\delta$, with $u_\delta^{\textrm{max}}=0.5$, $\alpha_\delta=1300000$ (cells/mm$^2$)$^2$. Here $u_r = u_K \equiv 0$.
        The units are $[C]=\textrm{cells/mm}^2, [t]=\textrm{h}$, and $u_r$ and $u_\delta$ are dimensionless.
        }
        \label{fig:exp_design_control_discrim_logistic_multiplicative}
\end{figure}

In contrast to the case of additive controls, where $u_\delta$ is ineffective in removing structural non-identifiability and $u_r$ and  $u_K$ are effective, as shown in Sec.~\ref{sec:results_iden}, in the multiplicative case $u_K$ is ineffective in removing structural non-identifiability, whereas $u_r,\ u_\delta$ are effective (Fig.~\ref{fig:exp_design_control_discrim_logistic_multiplicative}). For $u_r$, the strategy implied by the optimal control seems to be to turn the control on during most of the time span of the simulation, then ramp it down smoothly to zero as $t \to T$. This is because $u_r$ has an impact on the difference between the two solutions throughout the time span of the simulation, but towards the end of the simulation the cost of keeping it on outweighs the gain in the difference between model solutions. Comparing the shape of $u_r^*$ in Fig.~\ref{fig:exp_design_control_discrim_logistic_multiplicative}(a) to that in the additive control case shown in Fig.~\ref{fig:exp_design_control_discrim_logistic_ur}, we can observe that in both cases, $u_r^*$ has a local maximum near the inflection points of the two model solutions but, in the multiplicative case, $u_r^*$ decreases more rapidly after that point, whereas it stays near a constant value for a period of time in the additive case. This is because $u_r$ has a larger effect on the steady state of the models in the additive case compared to the multiplicative case. Therefore, as the model solutions approach the steady state $\keff$, the effect of $u_r$ on model difference diminishes in the multiplicative case, while the effect remains significant in the additive case as the model solution approaches the steady state.

In the case of $u_\delta$, the optimal control strategy involves setting its value to the maximum for most of the simulation time span, followed by a sharp decrease in the control as $t$ approaches $T$. This can be explained by the fact that the model solution $C_2$ is entirely unaffected by $u_\delta$, due to its parameter $\delta_2$ being equal to zero, whereas $C_1$ is influenced by $u_\delta$ at all time. Maintaining a high value of $u_\delta$ depresses $C_1$ relative to $C_2$, which is helpful for distinguishing the two model solutions in all stages of the experiment.

\subsection{Richards model with additive controls}
\label{sec:control_richards}

\begin{figure}[H]
    \centering
    \includegraphics[width=0.7\textwidth]{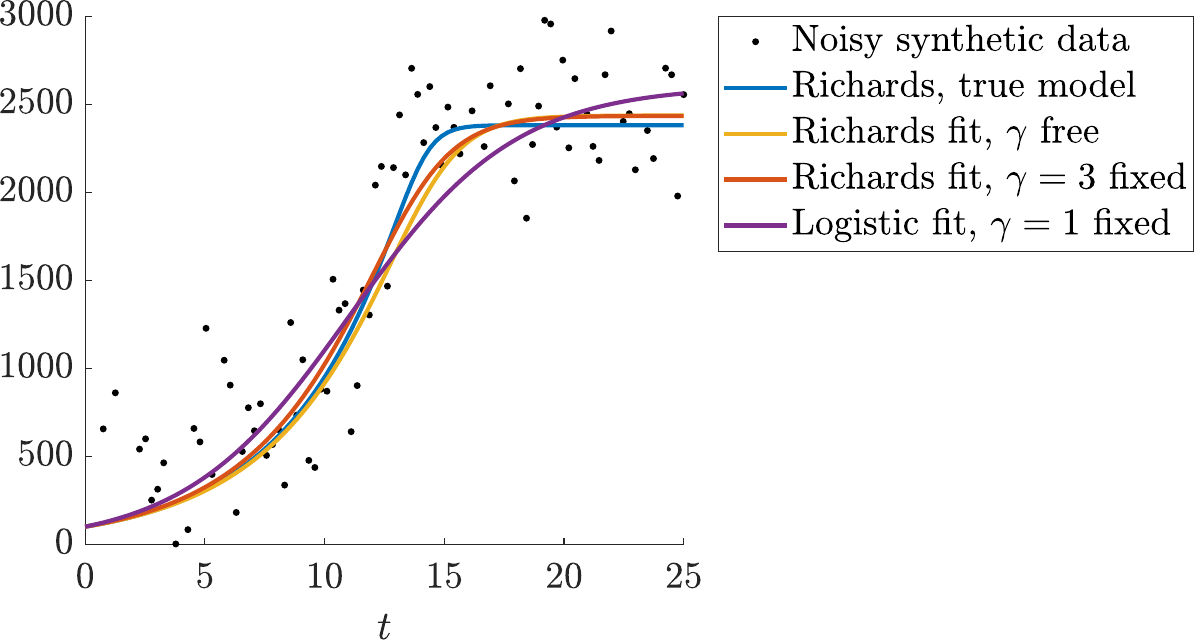}
    \caption{Illustrations of typical solutions of the logistic and Richards models. The blue, red, yellow, and purple curves are the solution to the Richards model (Eq.~\eqref{eqn:richards_basic}) with the parameters $(r,\gamma,K) = (0.225, 8, 2381), (0.235, 3, 2433), (0.222,3.709,2435),$ and $(0.291, 1, 2605)$, respectively. The units are $[r]=\textrm{h}^{-1}, [K]=\textrm{cell/mm}^2, [\gamma]=1$.}
    \label{fig:model_discrim}
\end{figure}

We provide another example for model discrimination using the Richards model, which is an extension of the logistic model for proliferation. Richards~\cite{richards1959FlexibleGrowthFunction} introduced an additional ``shape parameter" $\gamma$ to the logistic model, which provides more flexibility in describing the shape of the growth curve. The model can be written as
\begin{equation}
    \dv{C}{t} = rC\left[1- \left(\frac{C}{K}\right)^\gamma\right] = f_R(C; r, \gamma,K),
    \label{eqn:richards_basic}
\end{equation}
where the units and interpretations of the quantities in this model remain the same as in the logistic model. It has been shown in~\cite{simpson2022ParameterIdentifiabilityModel} that the Richards model, without control, is practically non-identifiable given a realistic amount of data, with the identifiability of the parameter $\gamma$ being especially poor. 
We plot typical solutions to the Richards model, and demonstrate its practical non-identifiability in Fig.~\ref{fig:model_discrim}. The black dots are synthetic data generated by perturbing the exact solution to the Richards model using a parameter set with $\gamma=8$ (blue curve), by adding Gaussian i.i.d. noise with $\sigma=400$ cell/mm$^2$. This parameter set and noise magnitude are within biologically realistic ranges, as found in~\cite{liu2024ParameterIdentifiabilityModel}. The other three curves are the solutions of the Richards model using parameter sets found by fitting the model to the data (i.e. MLEs), with $\gamma$ free (i.e. simultaneously fitting three parameters, $r, K$ and  $\gamma$), $\gamma=3$ or $\gamma=1$ fixed (i.e. fitting the remaining two parameters, $r$ and $K$), respectively. Notice that the differences between the curves are small, and the parameter set obtained by fitting the full model (corresponding to the yellow curve) is far from the true parameter values, a symptom of non-identifiability. If we mistakenly believed that the true parameter set has $\gamma=3$, or if the logistic model ($\gamma=1$) was the true model, the data would be insufficient to refute either belief, as the red and purple curves fit just as well.

Therefore, as an example, we seek to design an experiment to discriminate between the  two instances of the Richards model that correspond to the blue and red curve in Fig.~\ref{fig:model_discrim}, using additive control inputs:
\begin{subequations}
\label{eqn:default_params_control_discrim_richards}
\begin{align}
    \dv{C_1}{t} = (0.225+u_r(t))C_1\left[1- \left(\frac{C_1}{2381-u_K(t)}\right)^8\right] - u_\delta(t)C_1,\\
    \dv{C_2}{t} = (0.235+u_r(t))C_2\left[1- \left(\frac{C_2}{2433-u_K(t)}\right)^3\right]- u_\delta(t)C_2.
\end{align}
\end{subequations}

\begin{figure}[ht!]
     \centering
     \includegraphics[width=\textwidth]{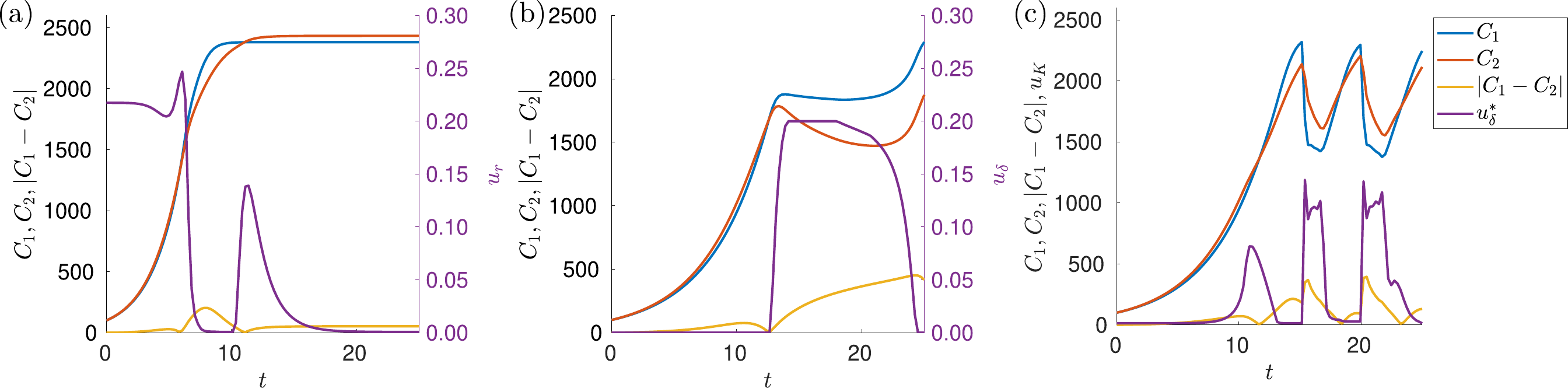}
        \caption[Optimal controls for model discrimination for Richards models]{Results of the FBS algorithm (Algorithm~\ref{alg:fbs}) applied to the optimal control problem for discriminating between Richards models with additive controls (Eq.~\eqref{eqn:optimal_control_discrimination_problem},~\eqref{eqn:default_params_control_discrim_richards}).
        (a) The control variable is $u_r$, with $u_r^{\textrm{max}}=0.4$ h$^{-1}$, $\alpha_r=30000$ h(cells/mm$^2$)$^2$. 
        (b) The control variable is $u_\delta$, with $u_\delta^{\textrm{max}}=0.2$ h$^{-1}$, $\alpha_\delta=500000$ h(cells/mm$^2$)$^2$. 
        (c) The control variable is $u_K$, with $u_K^{\textrm{max}}=1200$ cells/mm$^2$, $\alpha_K=0.03$  cells/mm$^2$. 
        The units are $[C]=\textrm{cells/mm}^2, [t]=\textrm{h}$, and $u_r, u_\delta$ and $u_K$ are dimensionless.
        }
        \label{fig:exp_design_control_discrim_richards}
\end{figure}

In Fig.~\ref{fig:exp_design_control_discrim_richards}, we present the optimal control obtained with the FBS algorithm in the case where one of the three control variables, $u_r$, $u_\delta$, and $u_K$, is applied individually. Unlike the case of distinguishing between logistic models in Sec.~\ref{sec:discrim_logistic_additive} and~\ref{sec:discrim_logistic_multiplicative}, all three controls are at least somewhat effective.
Out of the three, $u_r$ helps the least to distinguish between the two models. The optimal control $u_r^*$ never hits the imposed upper bound $u_r^\textrm{max}$, and it is turned on relatively briefly. The difference between the models in the presence of the control is negligibly greater than that in the absence of the control.

Both $u_\delta$ and $u_K$ are much more effective. The control strategy implied by $u_\delta^*$ allows both models to approach their respective carrying capacity, then turns $u_\delta$ on abruptly to maximum. This is a sensible strategy---the effect of $u_\delta$ is similar for both models in the earlier phase, where both model solutions, $C_1$ and $C_2$, display exponential growth, and $u_\delta$ lowers the effective growth rate $\reff$ by the same amount for both models, so it is useless to turn on the control in that phase. The control is turned on during the later saturation phase, where both model solutions approach their respective steady states. The steady state, with $u_\delta$ treated as a constant, can be written as
\[K\left(1-\frac{\delta+u_\delta}{r}\right)^{1/\gamma}.\]
A greater value of $\gamma$ diminishes the effect of $u_\delta$ on the steady state (since the exponent $1/\gamma$ approaches zero, so the steady state approaches $K$, effectively independently from $u_\delta$), therefore, the steady state of model 2, which has a lower $\gamma$, is decreased much more by a positive $u_\delta$ compared to that of model 1. The difference in steady states means turning the control on in the saturation phase allows the two models to be easily distinguished.

In contrast, $u_K$ modifies the $\keff$ of both models by the same amount, which means the case of using $u_K$ as the control variable requires a different strategy from the case of using $u_\delta$. The strategy in the case of $u_K$ is to first allow both models to get close, but not too close, to their respective carrying capacities, then cyclically and abruptly turn $u_K$ on and off so that both model solutions stay near, but not too close, to their respective carrying capacities. Recall that in the absence of the control, the two model solutions are best distinguished by their behaviour just before they approach $K$. This control strategy maximises the amount of time spent within the regime where the behaviours of the two models are most different, and thereby maximises the difference between the two model solutions.

\subsection{Logistic and Richards model with additive controls}

\begin{figure}[ht!]
     \centering
     \includegraphics[width=\textwidth]{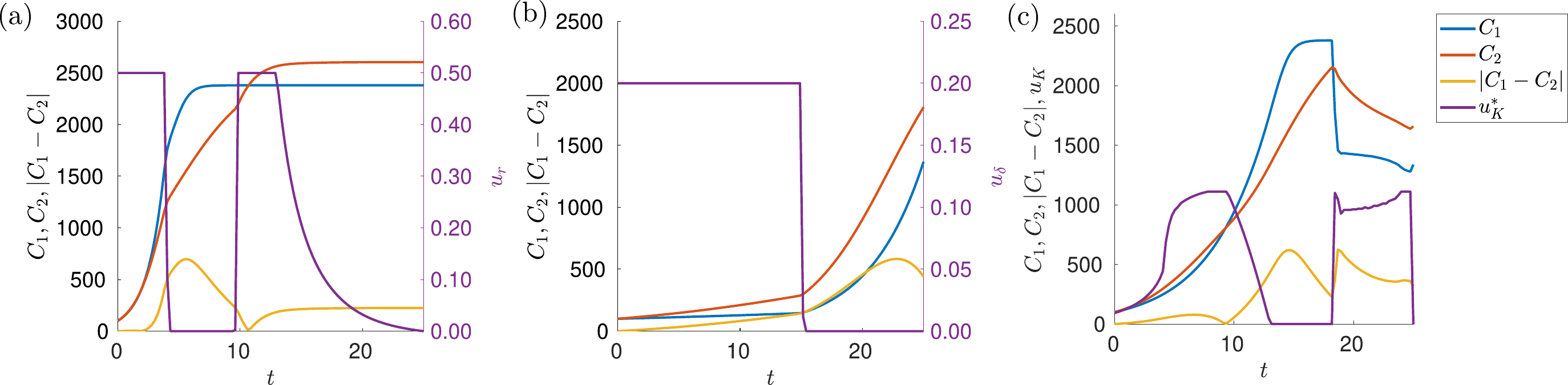}
        \caption{Results of the FBS algorithm (Algorithm~\ref{alg:fbs}) applied to the optimal control problem for discriminating logistic and Richards models (Eqs.~\eqref{eqn:optimal_control_discrimination_problem},~\eqref{eqn:default_params_control_discrim_richards_logistic}).
        (a) The control variable is $u_r$, with $u_r^{\textrm{max}}=0.5$ h$^{-1}$, $\alpha_r=30000$ h(cells/mm$^2$)$^2$. 
        (b) The control variable is $u_\delta$, with $u_\delta^{\textrm{max}}=0.2$ h$^{-1}$, $\alpha_\delta=30000$ h(cells/mm$^2$)$^2$. 
        (c) The control variable is $u_K$, with $u_K^{\textrm{max}}=1200$ cells/mm$^2$, $\alpha_K=0.03$  cells/mm$^2$. 
        The units are $[C]=\textrm{cells/mm}^2, [t]=\textrm{h}$, and $u_r, u_\delta$ and $u_K$ are dimensionless.
        }
        \label{fig:exp_design_control_discrim_richards_logistic}
\end{figure}

In the final example, we consider the problem of discriminating between instances of the logistic model and the Richards model, again using additive controls:
\begin{subequations}
\label{eqn:default_params_control_discrim_richards_logistic}
\begin{align}
    \dv{C_1}{t} & = (0.225+u_r(t))C_1\left[1- \left(\frac{C_1}{2381-u_K(t)}\right)^8\right] - u_\delta(t)C_1,\\
    \dv{C_2}{t} & = (0.291+u_r(t))C_2\left[1- \frac{C_2}{2605-u_K(t)}\right]- u_\delta(t)C_2.
\end{align}
\end{subequations}
In the absence of controls, the solutions to these two models are the blue and purple curves in Fig.~\ref{fig:model_discrim}, which are very similar. We plot the optimal controls for discriminating between these two models, as found by the FBS algorithm, in Fig.~\ref{fig:exp_design_control_discrim_richards_logistic}.

For the case of applying $u_r$, the strategy is to turn on the control in two stages. First, the control is turned on from the beginning of the experiment until both $C_1$ and $C_2$ are near their inflection point, at around $t\approx 5$ h, at which point the control is turned off. Since the Richards model approaches its carrying capacity much quicker than the logistic model after this point, we are able to observe  a significant difference between the two model solutions. Later, the control is turned on again from $t\approx 10$ h, and gradually ramped down to zero. Since $K_1$ and $K_2$ are different, control maximises the difference between the two models by allowing the logistic model to approach its carrying capacity quicker, so we can observe the two different limiting behaviours for longer.

The optimal control strategy for applying $u_\delta$ is to simply turn on the control to maximum limit from the beginning of the experiment until $t\approx 15$, then abruptly turn it off. The effective growth rate for the Richards model is more heavily impacted by the control than that of the logistic model, so turning on the control depresses $C_1$ more than $C_2$, allowing us to distinguish between the two models. The control is turned off in the later stages of the experiment as the difference between $C_1$ and $C_2$ is already significant, and the cost of applying the control in that stage outweighs the additional discrepancy it can cause between the two models. Lastly, for the case of applying $u_K$, the strategy is similar to the one seen in Fig.~\ref{fig:exp_design_control_discrim_richards}, where we use the control to keep both $C_1$  and $C_2$ away from their equilibrium values, so we can distinguish the two models by their behaviour as they approach their carrying capacity.

Summarising the results of this section, we find that it is possible to design an optimal control to enable us to differentiate between two models that were previously indistinguishable, or produce very similar solutions without such intervention. We can understand the optimal control strategies intuitively by looking at how the control variable affects the difference between the models at various points in time, using local sensitivities of the parameters associated with the control variable as a rough guide. Whether the control variables act on model parameters additively or multiplicatively can make a big difference, as an efficient control strategy in one scenario may prove entirely ineffective in the other.

\section{Discussion}\label{sec:exp_design_discussion}

This paper examined ways to optimally achieve two objectives, maximising the practical parameter identifiability of a model and discriminating between two models to the greatest extent, by designing a control input. This work extended existing research~\cite{faller2003SimulationMethodsOptimal,kreutz2009SystemsBiologyExperimental,litwin2022OptimalExperimentalDesign,steiert2012ExperimentalDesignParameter,vanlier2012BayesianApproachTargeted} by using profile likelihoods as a measurement of practical identifiability, and by considering a more general class of continuous control, which necessitates using a more sophisticated numerical algorithm to solve the optimal control problem, and investigated whether a continuous control can be more effective at achieving the objectives compared to simpler controls, such as bang-bang controls. 

We have established that the problems of model discrimination and improving identifiability, although related, represent distinct objectives and that, as a result, the optimal controls for achieving those objectives are different. This difference arises from the fact that the two objectives makes different assumptions on the availability of prior knowledge concerning the system. In tackling the problem of improving identifiability, we used an assumed set of parameter values to generate the synthetic data in order to evaluate the profile likelihoods. In contrast, for model discrimination, the model parameters are fixed at values that encode much more prior knowledge on the parameters. In the case where such knowledge is derived from an earlier experiment, in the model discrimination problem the optimal experimental design should ideally be very different from the earlier experiment to avoid collecting redundant information. On the other hand, there is less need to avoid repeating a similar experimental design if the goal is to maximise identifiability.

Our results highlight that structural non-identifiability issues can potentially be addressed by adding a control variable to the model, and incorporating the corresponding external stimulus into the experiment. The logistic model with a death term is structurally non-identifiable, as a change in $\delta$, the death rate, can be perfectly compensated for by a corresponding change in the growth rate, $r$, and the carrying capacity, $K$. Adding the appropriate control variable allows the system to be observed in a state that otherwise would not be accessible, providing additional information on the underlying mechanisms. For example, using $u_r$ as a control variable, with a window function for control, allows us to observe the system in two states, a state where the control is off, and another where the control is on. In this scenario, a change in $\delta$ can no longer be compensated for by changing $r$ and $K$ to obtain identical behaviour in both states. However, this approach can only succeed if knowledge of how the stimulus impacts the mechanisms (e.g.,~whether it induces an additive or multiplicative change in a parameter, and to what extent) is available. Without this knowledge, introducing the control variable introduces additional unknowns, which prevents the extraction of additional information on the parameters. For example, suppose we wish to introduce an external stimulus that modulates the growth rate, represented by the control variable $u_r$, but do not know how much this stimulus changes $r$ in advance. It would be necessary to introduce another unknown parameter, say $\rho$, as a coefficient of $u_r$ in the model. This $\rho$ must be estimated in the same way as the original model parameters $r, \delta$ and $K$. As a result, the inference problem becomes more complicated and requires more data, preventing any gains from introducing the control variable. One possibility of circumventing this issue is to test the control mechanism on a system with known dynamics. 

We also demonstrated that the optimal control for model discrimination depends on the control variable. The optimal strategy for control variables acting on different model parameters can be completely different. For example, in Fig.~\ref{fig:exp_design_control_discrim_richards}, the optimal strategies for the three control variables are qualitatively and quantitatively distinct. In some cases, the shape of the optimal continuous control is close to a window function, such as the case of an additive control $u_K$ for distinguishing two logistic models with death terms in Fig.~\ref{fig:exp_design_control_discrim_logistic_uk}, therefore it provides only a minor improvement upon the optimal bang-bang control. In this case, we might prefer to implement the bang-bang control in practice, since it requires relatively simple experimental techniques, whereas  continuous control requires the ability to modulate the control mechanism more precisely. In other cases, the optimal continuous control significantly out-performs the optimal bang-bang control, due to its increased flexibility. This is the case for an additive $u_r$  (Fig.~\ref{fig:exp_design_control_discrim_logistic_ur}), where the optimal continuous control has a significantly lower cost compared to the optimal bang-bang control. In these cases, the implementation of the continuous control is preferred, if it is feasible to carry out in the experiment. Note that bang-bang controls, or more generally piecewise constant controls, are more commonly used in control applications, due to the fact that they are both easier to compute and to implement compared to continuous controls. 

The FBS algorithm with adaptive update rate, based on PMP, is adequate for solving continuous optimal control problems, at least for the examples considered in this work. We have investigated alternative algorithms for solving the optimisation, such as direct control, or a hybrid method combining FBS and direct control in~\cite{dphil_thesis}. The results are given in Supplementary materials~\ref{apx:control_both}. In summary, the hybrid method can out-perform FBS in certain cases, but its implementation is more difficult. 

There are many future directions in which to extend the methods considered in this paper. We can generalise the methods in Sec.~\ref{sec:results_iden} by optimising the practical identifiability of multiple parameters at the same time, using multi-objective optimisation. It would also be useful to consider the case of applying multiple control inputs at the same time.
Finally, a key limitation of the approach proposed in this paper is the use of sets of fixed ``ground truth" parameter values to guide the design process. A more robust approach would be to replace the ground truth parameter sets with a prior distribution over the parameters, which would better account for uncertainty.

\section*{Statements}

\textbf{Data and code availability:} The code for the analysis in this paper is written in MATLAB, and will be provided on Github at \url{https://github.com/liuyue002/experiment_design}.

\bigskip\noindent
\textbf{Copyright:} For the purpose of open access, the authors have applied a CC BY public copyright licence to any author accepted manuscript arising from this submission.

\bigskip\noindent
\textbf{Contribution:} Conceptualization: YL, REB. Analysis: YL. Supervision: PKM, REB. Writing (original draft): YL. Writing (editing): YL, REB, PKM.

\bigskip\noindent
\textbf{Competing interests:} We declare no competing interests.

\bigskip\noindent
\textbf{Funding:} YL is supported by the Natural Sciences and Engineering Research Council of Canada (NSERC) through the Postgraduate Scholarships – Doctoral program, reference number PGSD3-535584-2019, as well as by the Canadian Centennial Scholarship Fund (CCSF). 

\newpage
\appendix

\section*{Supplementary materials}

\section{Forward-backward sweep  (FBS) algorithm}
\label{apx:pmp}
In this section, we summarise the FBS algorithm used in Sec.~\ref{sec:results_discrim}. This algorithm numerically finds solutions to the PMP problem. We adapt the version found in~\cite{sharpOptimalControlAcute2019}, adding a rule for adaptively reducing the update rate, $\omega$, during the iterations. More details on PMP and the FBS algorithm can be found in~\cite{dphil_thesis}.
\begin{algorithm}[H]
\caption{Forward-backward sweep with adaptive update rate}\label{alg:fbs}
\begin{enumerate}[label=(\Roman*)]
    \item Initialise $\bu^{0}(t)$
    \item Set $\omega \leftarrow \omega^0, J[\bu^0]\leftarrow\infty$
    \item For $i=1, 2,\ldots:$
    \begin{enumerate}[label=(\arabic*)]
        \item Solve the state equation, Eq.~\eqref{eqn:control_state_eqn}, with $\bu=\bu^{i}$ and $\bx(0)=\bx_0$, forward in time to obtain $\bx^{i}(t)$
        \item Solve the adjoint equation, 
        \begin{equation}
            \dv{\lambda_i}{t} = (-1)^i 2 g_1(\bx_{1}(t; \bu(t))) -g_2(\bx_{2}(t; \bu(t))) g_i'(x_i) - \lambda_i \pd{f_i}{x_i}, \label{eqn:control_adjoint_eqn}
        \end{equation}
        with $\bu=\bu^{i}$, $\blambda(T) =0$, backward in time to obtain $\blambda^{i}(t)$. 
        \item Compute the cost functional, $J[\bu^i]$, Eq.~\eqref{eqn:optimal_control_discrimination_problem}, using $\bu^i, \bx^i, \blambda^i$
        \item Check termination criterion: if $|J[\bu^i]-J[\bu^{i-1}]|<\epsilon$: return $\bu^i$ as the optimal control
        \item Set
        \begin{equation}
            \bunew(t) \leftarrow \argmax_{\bu' \in \mathcal{U}} H(\bx^i(t), \blambda^i(t); \bu'),
            \label{eqn:pmp_pointwise_optimality_unew}
        \end{equation}
        where $H(\bx, \blambda; \lambda_0, \bu) = -\Lag(\bx,\bu) + \blambda \cdot \bff(\bx,\bu)$
        \item Adapt update rate: if $J[\bu^i] > J[\bu^{i-1}]$: set $\omega \leftarrow \omega/2$
        \item Set $\bu^{i+1} \leftarrow (1-\omega)\bu^i + \omega \bu^{\textrm{new}}$
    \end{enumerate}
\end{enumerate}
\end{algorithm}

\section{Additional example for optimising parameter identifiability}

In this section, we provide an additional example for designing an experiment to optimise for parameter identifiability, using the same approach as in Sec.~\ref{sec:results_iden}. This is to demonstrate the flexibility of the methodology.
We consider a model of a hypothetical signalling pathway, which can also be interpreted as a regulatory network, consisting of three signalling proteins, named A, B, and C. In this model, A promotes both B and C directly, and B promotes C. This motif has been identified in several instances in biology, usually as a part of a more complex system, such as in the regulation of cancer-associated genes~\cite{awan2007RegulatoryNetworkMotifs}, transcriptional regulations in \textit{E. coli}~\cite{shen-orr2002NetworkMotifsTranscriptional}, and cytokine signalling~\cite{fruhbeck2005IntracellularSignallingPathways}.

The model equations are
\begin{equation}
    \dv{}{t} \begin{bmatrix}
        A\\B\\C
    \end{bmatrix} = \begin{bmatrix}
        -a_1&0&0 \\ b_1&-b_2&0 \\ c_1&c_2&-c_3 
    \end{bmatrix} \begin{bmatrix}
        A\\B\\C
    \end{bmatrix} + \begin{bmatrix}
        a_0 \\ b_0 \\ 0 
    \end{bmatrix} + \begin{bmatrix}
        u_a(t) \\ u_b(t) \\ u_c(t)
    \end{bmatrix}, \quad \begin{bmatrix}
        A(0)\\B(0)\\C(0)
    \end{bmatrix} = \begin{bmatrix}
        A_0\\ B_0\\ C_0.
    \end{bmatrix}
    \label{eqn:signalling_pathway}
\end{equation}
The state variables are $\bx=(A,B,C)$, the control inputs are $\bu=(u_a, u_b, u_c)$, and the model parameters are \[\btheta=(a_0, a_1, b_0, b_1, b_2, c_1, c_2, c_3, A_0, B_0, C_0).\]
All parameters are non-negative.
We assume that A and C, but not B, can be directly measured, giving the observation model
\begin{equation}
    g(\bx; \btheta) = \begin{bmatrix}
        A \\ C
    \end{bmatrix}.
    \label{eqn:signalling_pathway_observe}
\end{equation}

As this is a conceptual model, all quantities are considered to be non-dimensional.
In the absence of control, this model is structurally non-identifiable, since A can impact C both directly, and indirectly through B. Without measurements of B, it is difficult to distinguish between the two effects. It can be shown that the parameters $a_0, a_1, A_0, C_0$ are structurally identifiable, but the model on the whole is not (see Supplementary Materials~\ref{apx:structural}). The analysis also shows that there are a further five structurally identifiable combinations of parameters, but they are too cumbersome to be of any practical use.

For experimental design, we use the following ``ground truth" parameter values,
\begin{align}
    &(\hat{a}_0, \hat{a}_1, \hat{b}_0, \hat{b}_1, \hat{b}_2, \hat{c}_1, \hat{c}_2, \hat{c}_3, \hat{A}_0, \hat{B}_0, \hat{C}_0) = (1, \ 0.1, \ 0, \ 0.6, \ 0.4, \ 0.3, \ 0.4, \ 1, \ 1, \ 0, \ 0), \notag\\
    & T=50,\quad \Delta t=0.5, \quad \sigma=0.3.
\label{eqn:threethings_paramval}
\end{align}

\begin{figure}[ht!]
    \centering
    (a) No control\\
    \includegraphics[width=0.5\textwidth,valign=t]{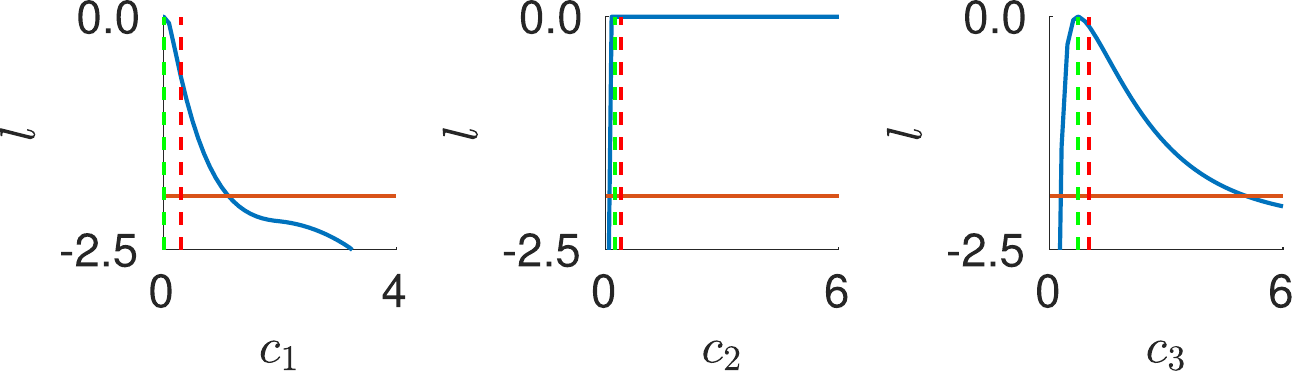}\\
    (b) $u_a$\\
    \includegraphics[width=0.5\textwidth,valign=t]{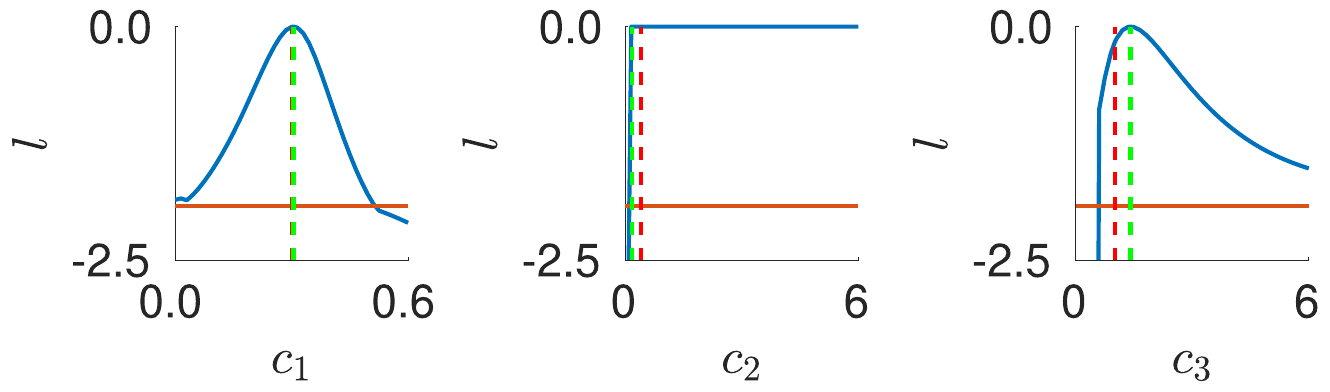}\\
    (c) $u_b$\\
    \includegraphics[width=0.5\textwidth,valign=t]{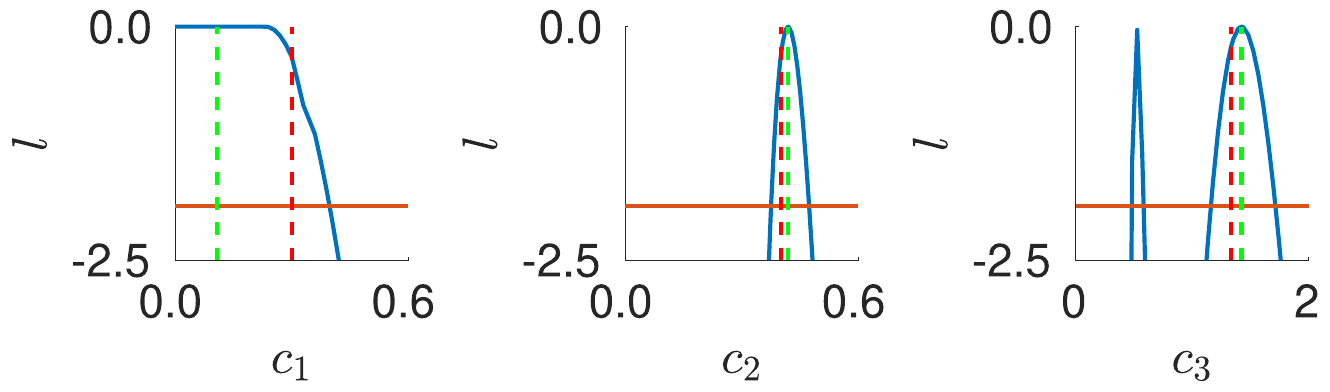}\\
    (d) $u_c$\\
    \includegraphics[width=0.5\textwidth,valign=t]{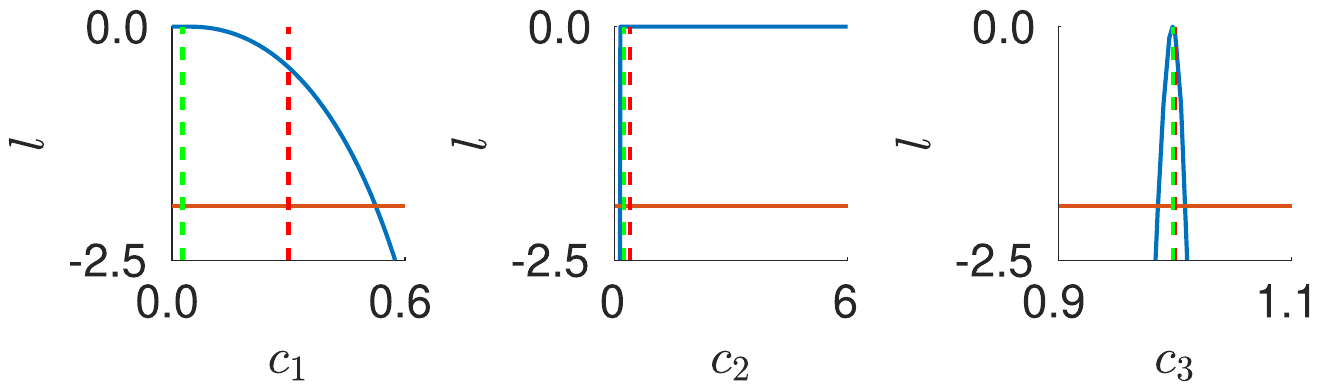}\\
    \caption{Profile likelihoods for the parameters $c_1, c_2$ and $c_3$ in the signalling pathway model (Eq.~\eqref{eqn:signalling_pathway}, \eqref{eqn:signalling_pathway_observe}), computed using synthetic datasets generated with the parameter values in Eq.~\eqref{eqn:threethings_paramval}. (a) No control is applied. (b-d) one of $u_a, u_b$ or $u_c$ is applied from $\tau_0=10$, for a period of $\tau=20$. The magnitudes of the controls are $u_{a,\textrm{max}}=2, u_{b,\textrm{max}}=18, u_{c,\textrm{max}}=18$, respectively. Note that the range of the $x$-axes varies between the plots.}
    \label{fig:threethings}
\end{figure}

Given the three possible control inputs $u_a, u_b, u_c$, for this simple demonstration, we consider applying one of them at a time. In Fig.~\ref{fig:threethings}, we present the profile likelihoods of the model parameters of interest when these controls are applied. For this cursory exploration, we choose, more or less arbitrarily, $\tau_0=10$ and $\tau=20$. We will discuss how to choose $\tau_0$ and $\tau$ appropriately in the next example.
The focus of this example is to demonstrate that the control inputs are not equally effective, and certain control inputs result in greater improvement in practical identifiability compared to others. 
We take $u_{a,\textrm{max}}=2, u_{b,\textrm{max}}=18, u_{c,\textrm{max}}=18$. Under these choices, the three possible control inputs elevate the equilibrium for $C$ by the same amount, which makes for a fair comparison.

Fig.~\ref{fig:threethings} shows how each of these controls impacts the practical identifiability of the model. Observe that when no control inputs are applied (Fig.~\ref{fig:threethings}(a)), the profile likelihood for $c_2$ appears perfectly flat. This indicates that this parameter is structurally non-identifiable, so a shift in the value of $c_2$ can be perfectly compensated for by an appropriate shift in the value of others, resulting in an identical model solution. Furthermore, while the parameters $c_1$ and $c_3$ appear to be practically identifiable, their confidence intervals are very wide. This would limit the usefulness of the estimates for these parameters. The MLE for $c_1$ is at zero, which misleadingly suggests that A does not influence C directly.

When $u_a$ is applied (Fig.~\ref{fig:threethings}(b)), the shape of the profile likelihood of $c_1$ changes so that it is now a roughly parabolic curve centered near $\hat{c}_1=0.3$. This parameter describes the rate at which $A$ up-regulates $C$, so it is unsurprising that manipulating the level of $A$ allows us to better pin down the value of $c_1$. The profile likelihoods of $c_2$ and $c_3$ remain qualitatively the same as in the no-control case (Fig.~\ref{fig:threethings}(a)). 

When $u_b$ is applied (Fig.~\ref{fig:threethings}(c)), all three parameters are practically identifiable. However, some identifiability issues remain. For example, the profile likelihood for $c_1$ is nearly perfectly flat for $0 \leq c_1 \leq 0.2$, so all values in that interval are equally valid as an estimate for $c_1$. Furthermore the profile likelihood for $c_3$ is bimodal, so the resulting confidence region is a union of two disjoint intervals, and we cannot tell which peak corresponds to the true value of the parameter. Despite these issues, the case of applying $u_b$ is the only case where all three parameters have finite confidence regions.

When $u_c$ is applied (Fig.~\ref{fig:threethings}(d)), $c_1$ and $c_3$ are identifiable, while $c_2$ is not. The main improvement compared to the no-control case (Fig.~\ref{fig:threethings}(d)) is that the confidence interval for $c_3$ is now much narrower. These results suggest that applying the control $u_b$ is the most effective for improving the identifiability of the model, as it is the only control input that results in all three parameters being practically identifiable. 
Since the value of $u_b$ is known to the experimenter, applying $u_b$ is akin to an indirect measurement of $B$. Since the non-observability of $B$ is the main source of non-identifiability in this model, it is not surprising that $u_b$ is the best choice for the control input. However, $u_a$ and $u_c$ would be better choices for a study focused on measuring the values of $c_1$ and $c_3$, respectively.

\section{Structural identifiability analysis}
\label{apx:structural}

In this section, we review the definition of structural identifiability analysis, and present the result of structural identifiability analysis for the models considered in this paper. The definitions given below follow~\cite{chapman2003StructuralIdentifiabilityNonlinear,chis2011StructuralIdentifiabilitySystems}.
An ODE model (Eq.~\eqref{eqn:gen_ode}) with an observation model (Eq.~\eqref{eqn:gen_observation}) is called
\begin{itemize}
    \item globally structurally identifiable, if for almost any $\btheta \in \Theta$, 
    \[\by(t; \btheta) = \by(t; \btheta') \ \forall \ t \Rightarrow \btheta = \btheta' ;\]
    \item locally structurally identifiable, if for almost any $\btheta \in \Theta$, there exists an $\epsilon>0$ such that
    \[\by(t; \btheta) = \by(t; \btheta') \ \forall \ t, \   \text{and } ||\btheta-\btheta'||<\epsilon \Rightarrow \btheta = \btheta' ;\]
    \item structurally non-identifiable, if it is not locally structurally identifiable.
\end{itemize}
Note that under this definition, globally structurally identifiable is a strictly stronger property than locally structurally identifiable.

\subsection{Logistic and Richards models}

We will use the Taylor series method~\cite{pohjanpalo1978SystemIdentifiabilityBased} to perform structural identifiability analysis for the logistic and Richards models, which can be summarised as a single equation,
\begin{equation}
\dv{C}{t} = rC\left[1-\left(\frac{C}{K}\right)^\gamma \right] - \delta C.\label{eqn:logistic_richards_general}
\end{equation}
The logistic model has $\gamma=1$, and the Richards model has $\gamma > 0$. We will use a prime ($(\cdot) \ '$) to denote a derivative with respect to time in this section. 
The Taylor series approach for structural identifiability analysis assumes that the solution is smooth and can be expressed as a Taylor series in terms of time, and certain regularity conditions on the right-hand side of the ODE model holds. Under these assumptions,~\cite{pohjanpalo1978SystemIdentifiabilityBased} showed that the model is structurally identifiable if the parameters can be uniquely determined if measurements of the system and all its derivatives in time at $t=0$ are given, that is, $C(0)$ and $C^{(k)}$ for $k \geq 1$. Knowing the values of these derivatives is equivalent to knowing the coefficients of the Taylor series of the model solution at $t=0$. 
In practice,  we attempt to write the model parameters (or their combinations) as functions of $C(0)$, $C'(0)$, and the higher order derivatives, which are considered known quantities. If a parameter, or a parameter combination, can be written entirely in terms of known quantities, then they are at least locally structurally identifiable. If a parameter or parameter combination cannot be expressed in terms of known quantities alone, then they are structurally non-identifiable.

The leading order derivatives of the model solution can be written as:
\begin{subequations}
\begin{align}
    C(0) &= C_0 \label{eqn:structural_taylor0}\\
    C'(0) &= r C_0 (1-(C_0/K)^\gamma) - \delta C_0 \label{eqn:structural_taylor1}\\
    C''(0) &= C'(0) [r-r(1+\gamma) (C_0/K)^\gamma-\delta] \label{eqn:structural_taylor2}\\
    C'''(0) &= C''(0)^2/C'(0)- C'(0)^2 [r(\gamma+1)\gamma C_0^{\gamma-1}/K^{\gamma}]\label{eqn:structural_taylor3} \\
    C''''(0)&= \begin{cases}
        A - C'(0)^2 [r(\gamma+1)\gamma(\gamma-1)C_0^{\gamma-2}/K^{\gamma}] \text{\ if \ } \gamma \neq 1\\
        A \hphantom{ - C'(0)^2 [r(\gamma+1)\gamma(\gamma-1)C_0^{\gamma-2}/K^{\gamma}]} \text{\ \ \ if \ } \gamma=1
    \end{cases},\label{eqn:structural_taylor4}
\end{align}
\end{subequations}
where $A$ denotes terms that can be written exclusively using known quantities. For both the Richards and logistic models, Eq.~\eqref{eqn:structural_taylor0} shows that the initial condition, $C_0$, can be determined from data, if it weren't already known. 

For the Richards model, Eq.~\eqref{eqn:structural_taylor3} and Eq.~\eqref{eqn:structural_taylor4} together provide two polynomial constraints on $\gamma$ and the parameter combination $r/K^\gamma$, which means that in general, there are at most a finite number of solutions for both, so $\gamma$ and $r/K^\gamma$ are at least locally identifiable. Given values for $\gamma$ and $r/K^\gamma$, Eq.~\eqref{eqn:structural_taylor2} allows $r-\delta$ to be uniquely identified, so $r-\delta$ is at least locally identifiable. However, the parameters $r, K, \delta$ cannot be identified individually, since $r$ always appears in the combinations of either $r-\delta$ or $r/K^\gamma$ in Eq.~\eqref{eqn:structural_taylor1} or the higher derivatives, so the Richards model is overall non-identifiable. If the explicit death term is not present, or equivalently, $\delta=0$ is fixed, then all parameters are at least locally identifiable.

For the logistic model, which has $\gamma=1$ fixed, Eq.~\eqref{eqn:structural_taylor3} shows that the parameter combination $r/K$ can be determined from data, which together with Eq.~\eqref{eqn:structural_taylor2}, allows $r-\delta$ to be uniquely determined. None of the three parameters can be uniquely determined individually, for the same reason as for the Richards model. 
Therefore, the logistic model is structurally non-identifiable. However, knowing the value of one of $r, \delta, K$ will allow the other two parameters to be uniquely identified. Therefore, the logistic models without the explicit death term (Eq.~\eqref{eqn:logistic_richards_general} with $\gamma=1$ and $\delta=0$ fixed) is globally identifiable.

\subsection{Signalling pathway model}

We again use the Taylor series method to perform structurally identifiability analysis for the signalling pathway model (Eq.~\eqref{eqn:signalling_pathway}, \eqref{eqn:signalling_pathway_observe}).
It has been shown that, for linear models such as the signalling pathway model, we need to compute the Taylor series to at most $2n_x-1$ degrees to determine structural identifiability~\cite{vajda1984StructuralIdentifiabilityLinear}, where $n_x$ is the number of state variables. For the model at hand, this means we need to consider up to order 5.

The constant term in the Taylor series for the observable state variables $A, C$ are $A_0$ and $C_0$, respectively, therefore these two parameters are structurally identifiable. The coefficients of the Taylor series for $A(t)$ are
\begin{subequations}
\begin{align}
    A'(0) &= -a_1 + a_0, \label{eqn:structural_threethings_A1}\\
    A''(0) &= a_1^2 -a_0 a_1 ,\label{eqn:structural_threethings_A2}\\
    \dv{^k}{t^k}A(0) &= (-1)^n \left[ a_1^k - a_1^{(k-1)}a_0 \right].\label{eqn:structural_threethings_Ak}
\end{align}
\end{subequations}
Eqs.~\eqref{eqn:structural_threethings_A1} and \eqref{eqn:structural_threethings_A2} imply that there are at most two possibilities for $a_0$ and $a_1$. Under conditions of genericity, one of these possibilities can be eliminated using Eq.~\eqref{eqn:structural_threethings_Ak}, therefore $a_0$ and $a_1$ are at least structurally locally identifiable, and are globally identifiable in generic cases.

The first two coefficients of the Taylor series for $C(t)$ are
\begin{subequations}
\begin{align}
    C'(0) &= c_2 B_0 + c_1, \label{eqn:structural_threethings_C1}\\
    C''(0) &= c_1(-a_1+a_0-c_3) + c_2 (b_1+b_0)-(b_2+c_3) c_2 B_0,\label{eqn:structural_threethings_C2}
\end{align}
\end{subequations}
and the coefficients of the higher order terms are increasingly complicated. Each of these coefficients, up to the fifth one, represents a structurally identifiable combination of parameters. Since there are seven remaining parameters $(b_0, b_1, b_2, c_1, c_2, c_3, B_0)$, which is more than the number of constraints (five), the model is structurally non-identifiable.

\section{Direct control as an alternative to PMP}
\label{apx:control_both}

In this section, we discuss direct control as an alternative to PMP for the purpose of optimally designing an experiment for discriminating between two models. Direct control, in this context, means solving an infinite-dimensional functional optimisation problem by applying a discretisation scheme, resulting in  a discrete-time problem involving only finitely many variables, that approximates the original problem. This finite dimensional problem is then solved using a general-purpose optimisation algorithm.
We discretise the time interval $[0,T]$ into $n_t$ evenly spaced points $t_i=i \Delta t$, $i=1, \dots, n_t$, where $\Delta t = T/n_t$. The control $u(t)$ is parameterised as its values at $t_i$, interpolated linearly between these values, where we choose $n_t=100$. We use \textit{fmincon} with the default interior point algorithm~\cite{byrd1999InteriorPointAlgorithm}, with a termination criterion based on the Karush-Kuhn-Tucker (KKT) first-order optimality measure, which is akin to the first derivative condition in the presence of constraints. This is a reasonable choice, as the interior-point algorithm was found to be suitable for directly optimising a variety of control problems in~\cite{laurent-varin2007InteriorPointApproachTrajectory}. We initialise the optimisation at $u(t) \equiv 0$ when applying direct optimisation alone. 

The caveat of direct optimisation is that a general-purpose algorithm cannot exploit the structure of the original control problem. We also need a sufficiently large  $n_t$ for the discrete-time system to be able to approximate the original continuous-time system well, which means that the discrete-time optimisation problem is very high-dimensional, and we expect the optimisation algorithm to converge slowly, and to be sensitive to the choice of initial point where we start our search. To partially mitigate these problems with direct optimisation, we also consider a hybrid method, where we initialise the aforementioned direct optimisation procedure at the output of the FBS algorithm.

\begin{figure}[ht!]
    \centering
    \includegraphics[width=0.9\textwidth]{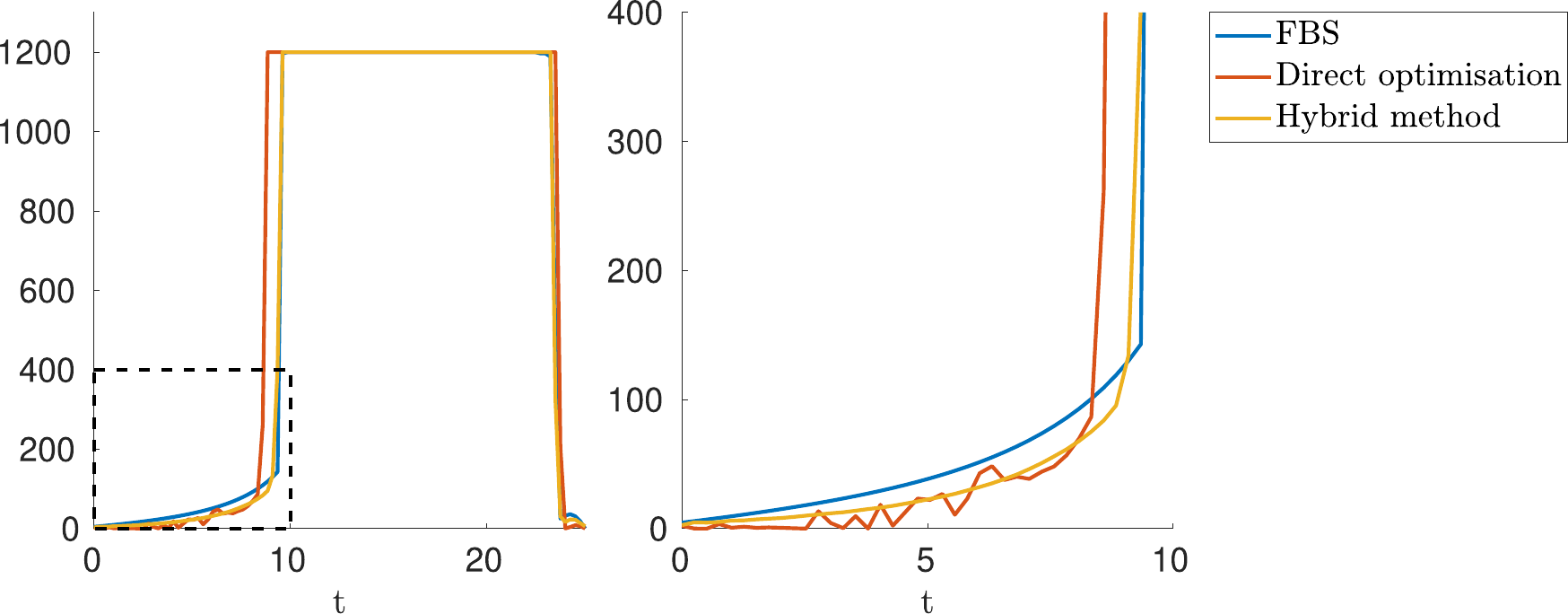}
    \caption[Comparison of numerical algorithms for optimal control problem]{Results from applying the FBS algorithm, direct optimisation, and the hybrid method to the optimal control problem in Eq.~\eqref{eqn:optimal_control_discrimination_problem}. The plot on the right is a zoomed-in version of the area inside the dashed rectangle in the plot on the left, where the three solutions noticeably differ.
    The control found by the FBS algorithm is the same as the one presented in Fig.~\ref{fig:exp_design_control_discrim_logistic_uk}.
    The units are $[t]=\textrm{h}, [u_K]=\textrm{cells/mm}^2$.
    }
    \label{fig:exp_design_control_discrim_logistic_uk_method_compare}
\end{figure}

To illustrate these approaches, we revisit the problem of discriminating between two logistic models in Eq.~\eqref{eqn:logistic_discrimination_additive}.
The outputs of direct optimisation and the hybrid methods  are given in Fig.~\ref{fig:exp_design_control_discrim_logistic_uk_method_compare}, along with the output of the FBS algorithm, originally given in Fig.~\ref{fig:exp_design_control_discrim_logistic_uk}. Notice that the optimal controls given by all three methods are very close. There are some noticeable numerical artifacts in the form of oscillations in the output of the direct optimisation procedure. The FBS algorithm terminated after 73 iterations, and produced a control with the objective value of $J=-784470$ (cells/mm$^2$)$^2$. Direct optimisation terminated after 991 iterations of \textit{fmincon}, although each of its iterations, which only require evaluating the objective function $J$ and its partial derivatives with respect to the discretised control variables, is much faster than one iteration of FBS. Its output achieved an objective value of $J=-782476$ (cells/mm$^2$)$^2$, which is close to, but a little worse, than that of the FBS algorithm. Finally, the hybrid method required 67 iterations of \textit{fmincon} (in addition to the computation of the FBS algorithm), and achieved an objective value of $J=-785081$ (cells/mm$^2$)$^2$, the best out of the three methods.

This example show that the FBS algorithm should be preferred over the direct optimisation algorithm with naive or uninformed initialisation. However, direct optimisation can be initialised at the output of the FBS algorithm to further refine the control if desired.


\printbibliography


\end{document}